\documentclass[preprint]{aastex63}

\usepackage{mathtools}
\usepackage{amssymb}
\usepackage{amsmath}
\usepackage{glossaries}

\makeglossaries
\glsunsetall
\newcommand{\zs}{z_{\rm s}}
\newcommand{\zd}{z_{\rm d}}
\newcommand{\vein}{V_{\rm Ein}}
\newcommand{\rein}{\theta_{\rm Ein}}
\newcommand{\mbh}{M_{\rm BH}}
\newcommand{\fq}{f_{\rm QL}}
\newcommand{\avfq}{\langle\fq\rangle}
\newcommand{\nql}{n_{\rm QL}}
\newcommand{\mlim}{m_{\rm lim}}
\newcommand{\flensed}{f_{\rm lensed}}
\newcommand{\dd}{\mathrm{d}}
\newcommand{\msun}{M_{\rm sun}}

\defcitealias{Courbin+10}{C10}
\defcitealias{Meyer+19}{M19}
\defcitealias{Oguri+10}{OM10}
\defcitealias{Collett15}{C15}

\revised{\today}

\shorttitle{HULQ I. Number Estimates of QSO Lenses}
\shortauthors{Taak \& Im}

\begin{document}

\title{High-$z$ Universe probed via Lensing by QSOs (HULQ) I. \\
Number Estimates of QSO-QSO and QSO-Galaxy Lenses}

\email{yctaak@astro.snu.ac.kr, mim@astro.snu.ac.kr}

\author{Yoon Chan Taak}
\affiliation{Center for the Exploration of the Origin of the Universe (CEOU)}
\affiliation{Astronomy Program, Department of Physics \& Astronomy, Seoul National University, 1 Gwanak-ro, Gwanak-gu, Seoul 08826, Korea}

\author{Myungshin Im}
\affiliation{Center for the Exploration of the Origin of the Universe (CEOU)}
\affiliation{Astronomy Program, Department of Physics \& Astronomy, Seoul National University, 1 Gwanak-ro, Gwanak-gu, Seoul 08826, Korea}

\begin{abstract}
It is unclear how galaxies and their central supermassive black holes (SMBHs) co-evolve across cosmic time, especially for the non-local universe ($z \gtrsim 0.5$). The High-$z$ Universe probed via Lensing by QSOs (HULQ) project proposes to utilize quasi-stellar object (QSO) host galaxies acting as gravitational lenses (QSO lenses) to investigate this topic. This paper focuses on the feasibility of this project, i.e., whether sufficiently large numbers of QSO lenses are expected to be found in various concurrent and future imaging surveys. We find that $\sim 440$ QSO lenses will reside in the Hyper Suprime-Cam Wide survey (HSC/Wide), which is expected to be the most prolific concurrent survey, with this number being boosted by one to two orders of magnitude (to $\sim 10000$) with upcoming surveys such as that conducted with the Large Synoptic Survey Telescope (LSST). We discuss several methods of how to study the redshift evolution of the $\mbh - \sigma_*$ relation, which is a stand-out illustration of the co-evolution. In addition, we demonstrate how the intimacy of lensed images to the bright deflector QSO for most systems will affect the \textit{detectability} of QSO lenses. We estimate that only $\sim 82$ and 900 will be detectable for HSC/Wide and LSST, respectively; the decrease is significant yet still yields an acceptable sample for the main objective. This decrease will be less of a problem for space-based imaging surveys, for their small point spread function FWHMs will allow detections of lensed images lying relatively close to the deflector QSO, and thus unveil the less massive yet more numerous QSO hosts.
\end{abstract}


\section{Introduction} \label{sec:intro}
Supermassive black holes (SMBHs) are known to reside in the centers of almost all galaxies \citep{Kormendy+95, Magorrian+98, Gebhardt+00a}. A significant portion of these SMBHs are observed as quasi-stellar objects (QSOs), which are among the brightest objects in the universe. Their high luminosities enable us to discover them at all post-reionization redshifts \citep{Schneider+02,Schneider+10,Paris+17,Paris+18}, even those in the early universe \citep{FanX+06,KimY+15,KimY+19,Banados+16,JiangL+16,Matsuoka+16,JeonY+16,JeonY+17,WangF+17,ShinS+20}. 

Masses of a considerable number of these SMBHs ($\mbh$) have been measured via multiple approaches; gas and stellar kinematics around the SMBH are used mainly for nearby, resolved galaxies (e.g., \citealt{Kormendy+95,Ferrarese+00,Gebhardt+00a}), and reverberation mapping is the primary method for distant QSOs \citep{Blandford+82,Peterson+04,Bentz+09,ShenY+15a}. Based on the broad line region size $-$ luminosity ($R-L$) relation \citep{Wandel+99,Kaspi+00,Kaspi+05} from reverberation mapping studies, and assuming virial equilibrium, $\mbh$ can be estimated with a single spectrum; this technique is commonly known as the single epoch method \citep{Woltjer59,Vestergaard+06,Vestergaard+09,KimD+10,KimD+15,Jun+15}.

There are well-established relations between $\mbh$ and properties of the bulge component of galaxies, among which the most famous is the $\mbh - \sigma_*$ relation \citep{Ferrarese+00, Gebhardt+00a, Kormendy+13}, where $\sigma_*$ denotes the stellar velocity dispersion of the bulge. The tightness of such correlations imply that SMBHs play an important role in galaxy evolution \citep{Woo+10}, but details remain veiled (\citealt{PengC07,Jahnke+11} propose a different scenario). 

To explore how this so-called ``co-evolution'' actually took place, it is crucial to examine the evolution of the forementioned correlations, if at all. Unfortunately, this is especially difficult at high redshifts, starting from $z \sim 1$, where decomposition of galaxies from their central QSOs becomes a delicate task due to their small angular sizes \citep{PengC+06,Decarli+10,Merloni+10}. This turns out to be more complicated due to the utterly strong luminosities of QSOs causing them to outshine their host galaxies even for closer targets \citep{Gebhardt+00b}, which makes detecting and studying the host galaxies a challenging task. 

In the case that these QSO host galaxies act as strong gravitational lenses (hereafter deflectors, to avoid confusion with \textit{lens systems}) and distort the shapes of sources behind, the masses of these QSO hosts can be estimated using gravitational lensing (GL). The galaxies are much heavier than their hosted QSOs, albeit being less luminous, and any sort of GL features around QSOs will be predominantly due to the mass of the galaxies. Consequently, careful modeling of these GL features will allow us to measure the masses of the QSO host galaxies to high accuracy.

As mentioned above, QSOs are discovered at low to high redshifts, and their $\mbh$ are relatively easy to determine with a single spectrum. Therefore, QSO host galaxies that act as GL deflectors (hereafter QSO lenses) yield a unique method of studying galaxies and their central SMBHs. Furthermore, a sufficiently large sample of QSO lenses at various redshifts will allow us to examine the co-evolution of SMBHs and their host galaxies. 

In addition, galaxy-scale GL systems have been used to constrain the density profiles of galaxies, in joint analyses with stellar kinematics or weak lensing \citep{Koopmans+06,Gavazzi+07}. QSO lenses will provide knowledge about the mass profiles of QSO host galaxies, and allow us to inspect whether QSO hosts are different from normal galaxies with quiescent SMBHs. 

There has only been one group studying QSO lenses so far; the authors use Sloan Digital Sky Survey (SDSS) spectra of QSOs to look for emission lines originating from sources at redshifts further than those of the QSOs themselves. The first QSO lens, found from SDSS Data Release 7 (DR7) QSO spectra \citep{Schneider+10}, was presented by \cite{Courbin+10} (hereafter \citetalias{Courbin+10}). \cite{Courbin+12} confirmed three additional samples (while rejecting the first with \textit{Hubble Space Telescope} (\textit{HST}) imaging data), and \cite{Meyer+19} (hereafter \citetalias{Meyer+19}) have released a list of QSO lens candidates with additional spectra from DR12 \citep{Paris+17}. These studies, however, have the following limitations. 

First, the redshifted emission lines from the lensed image must be detected using SDSS spectra. Since strong and multiple emission lines are required, the survey relies heavily on hydrogen lines in the optical, such as H$\alpha$ and H$\beta$. The wavelength coverage of SDSS implies that the source, and in turn the deflector, must be located at relatively low redshifts. The furthest confirmed QSO lens is only at $z \approx 0.3$, which is insufficient for high redshift BH-galaxy co-evolution studies. Second, the use of the SDSS spectra also indicates that the lensed images must be located within the spectroscopy fiber, which has diameters of 3$\arcsec$ prior to the Baryonic Oscillation Spectroscopy Survey (BOSS), and 2$\arcsec$ for BOSS and later surveys \citep{Dawson+13}. This size is comparable to the typical Einstein radii of galaxy-scale lens systems, within which the source must be located in for it to be strongly lensed, so the SDSS fiber may not contain the lensed images for some systems. Third, \citetalias{Meyer+19} have published only a list of candidates, and have not confirmed the QSO lenses yet. Since no follow-up imaging observations have been conducted, the deflector-image configurations cannot be confirmed. Thus, these candidates require spectroscopic data for confirmation, and additional imaging data for further studies regarding the QSO hosts. Finally, in the case that the source has no strong emission lines, e.g., elliptical galaxies, it is near impossible to detect these lens systems with spectroscopy.

Searching for QSO lenses in imaging data can relieve all four constraints to some extent. First, the dependence on the optical wavelengths is removed, and sources at high redshifts that are bright in the restframe ultraviolet can be detected from optical imaging data, thus allowing the deflector QSOs to be located at higher redshifts also. The second issue is trivially taken care of since there is no fiber, and the entire neighborhood around the QSO can be scanned for possible lensed features. The third problem is also alleviated, since the configurations of the lens systems are known to a certain extent, although lacking in redshift information. This implies that follow-up spectroscopy is crucial in confirmation of these systems. Fourth, imaging data does not discriminate the type of sources; regardless of the existence of emission lines, they can be detected as long as they are sufficiently bright within the wavelengh range of the filter in question. Thus, QSO lens surveys are more promising when using imaging data. 

The issue, however, is that previous imaging surveys have been insufficient for a large sample of QSO lenses, either lacking in depth or area. Recently, several surveys with larger sizes and deeper limiting magnitudes have been undertaken; the Hyper Suprime-Cam Survey Strategic Program (HSC-SSP) Wide Survey (HSC/Wide; \citealt{Aihara+18a}), the Panoramic Survey Telescope and Rapid Response System 1 (Pan-STARRS1) 3$\pi$ Steradian Survey (PS1/3$\pi$; \citealt{Chambers+16}), the Kilo-Degree Survey (KiDS; \citealt{deJong+13}), and the planned Large Synoptic Survey Telescope (LSST) surveys \citep{Ivezic+19} are some examples. Furthermore, a number of space missions, such as \textit{Euclid} \citep{Laureijs+11}\footnote{This paper is outdated, so the \textit{Euclid} Consortium webpage (https://www.euclid-ec.org) is used for updated details.}, the \textit{Chinese Space Station Telescope (CSST)}\footnote{Currently, there is no overview paper for \textit{CSST}, so details of the Chinese Space Station Optical Survey (CSS-OS) are from \cite{GongY+19}.}, and the \textit{Wide Field InfraRed Survey Telescope (WFIRST)}\footnote{Currently, there is no overview paper for \textit{WFIRST} surveys, so details of the \textit{WFIRST} Wide Field Instrument High-Latitude Survey (\textit{WFIRST}/WFIHLS) are from the \textit{WFIRST} webpage (https://stsci.edu/wfirst).} are proposed to be launched within the decade, and are expected to carry out imaging surveys with no atmospheric seeing, meaning their image qualities will be incomparable to ground-based surveys. These surveys have the following characteristics: deep limiting magnitudes ($\gtrsim 23$ mag) and high image qualities (pixel scale $\&$ seeing $\lesssim 1\arcsec$) to enable the detection of the faint lensed images separated from the deflector QSO, and large areas ($\gtrsim$ 100 deg$^2$) that are required to compensate for both the low number density of QSOs and the intrinsically low probability of galaxies acting as deflectors. Therefore, a survey for QSO lenses is likely conceivable nowadays. 

The objective of the High-$z$ Universe probed via Lensing by QSOs (HULQ) project is to search for QSO lenses at various redshifts to explore the evolution of SMBH-galaxy correlations. To check the feasibility of this project, this paper aims to estimate the number of QSO lenses that are detectable in various recent and upcoming surveys. Future papers will discuss new findings of QSO lenses, and how these discoveries will affect our understanding of the co-evolution of SMBHs and their host galaxies.

This paper is organized as follows. Section 2 demonstrates how the number of QSO lenses is calculated, and the data used in these calculations are described in Section 3. Results are shown in Section 4, and further issues, such as factors that may affect our estimates, are discussed in Section 5. Section 6 summarizes our work.

The AB magnitude system is used throughout this paper \citep{Oke+83}. We use a standard $\Lambda$CDM cosmology model with $H_0 = 70$ km s$^{-1}$ Mpc$^{-1}$, $\Omega_{\rm M} = 0.3$, and $\Omega_\Lambda = 0.7$, which is supported by recent observational studies \citep{Im+97,Planck+18}.

\section{Methodology} \label{sec:method}

The number of QSO lenses can be calculated following the subsequent arguments \citep{Dobler+08}. We define the Einstein volume ($\vein$) of a deflector to be the region within which a source must be located to be lensed by the deflector. The probability of a specific object to act as a deflector is equivalent to the number of sources that are lensed to be brighter than the depth of the image and located within $\vein$ for that deflector, which in turn is equal to the probability that a sufficiently bright source is within $\vein$ of the deflector, summed for all possible sources. 

Expressing this in equation form, for an object at redshift $\zd$ with velocity dispersion $\sigma$, the probability that this object is a deflector to a background source, $\fq$, is 
\begin{equation}
\begin{aligned}
\fq&=\int_{\vein} n_{\rm s}(\zs, \: L_{\rm lim}/\mu_2) \: \dd V\\
&=\int_{\zs=\zd}^{\infty} \dfrac{\dd V}{\dd \zs \: \dd\boldsymbol{\vec{u}}} \int_{\rm mult} n_{\rm s} (\zs, \: L_{\rm lim}/\mu_2) \: \dd \boldsymbol{\vec{u}} \: \dd \zs, \label{eq:prob}\\ 
\end{aligned}
\end{equation}
where $\zs$ is the redshift of the source, $L_{\rm lim}$ is the limiting luminosity of the imaging data that is being searched, $\mu_2$ is the magnification of the second brightest image, $n_{\rm s}(\zs, \: L_{\rm lim}/\mu_2)$ is the number density of sources at $\zs$ detectable as lensed images with the second brightest image brighter than $L_{\rm lim}$, $\boldsymbol{\vec{u}}$ is the angular position vector from the deflector in the source plane, and $\int_{\rm mult}$ indicates that this integral is calculated for the multiply-imaged region only, i.e., within $\vein$. The first component depends on the cosmology, and following \cite{Hogg99}, can be expressed as
\begin{equation}
\begin{aligned}
\dfrac{\dd V}{\dd \zs \: \dd \boldsymbol{\vec{u}}} = \dfrac{c}{H_0} \: \dfrac{(1+\zs)^2 \: D_{\rm os}^2}{[\Omega_{\rm M} \: (1+\zs)^3 \: + \: \Omega_\Lambda]^{1/2}} \equiv F(\zs),
\end{aligned}
\end{equation}
where $D_{\rm os}$ is the angular diameter distance from the observer to the source redshift, so the entire component is simply a function of $\zs$. To compute the multiply-imaged region, we need to know the parameters $\zd$ and $\sigma$ of the plausible deflectors, and to obtain $n_{\rm s}$ and $\zs$, it is necessary to determine which samples can be used as likely sources. Therefore, to calculate the above equation, we need to constrain the deflector and source populations. 

Since the main subject of this study is QSO lenses, it is natural to use QSO host galaxies as the deflector sample. Thus, we use the velocity dispersion function (VDF) of QSO host galaxies as a function of redshift ($\zd$), which is explained in Section \ref{subsec:deflector}.
For the source population, the source luminosity functions (LFs) with respect to redshift are required. We assume that QSOs and galaxies are the predominant sources; these are the only samples with sufficient luminosities and number densities. These two populations are described in detail in Section \ref{subsec:source}.

To calculate $n_{\rm s}$, we integrate the LFs over luminosities brighter than the flux limit at a specific $\zs$. As shown in Eq. \ref{eq:prob}, the flux limit is altered by $\mu_2$, the magnification of the second brightest image; this is because multiple images (more than two) are required to confirm the object as a lens system. It is possible for a source to simply be distorted into only one image and still be confirmed as a lens system, but we assume that all sources are points, and ignore this possibility. Assuming singular isothermal spherical mass distributions, the multiply-imaged region is within the Einstein radius ($\rein$), expressed as 
\begin{equation}
\begin{aligned}
\rein= &\:4\pi \: \Big( \dfrac{\sigma}{c} \Big)^2 \: \dfrac{D_{\rm ds}}{D_{\rm os}}, \label{eq:rein}\\
\end{aligned}
\end{equation}
where $D_{\rm ds}$ is the angular diameter distance from the deflector redshift to the source redshift, so $\rein$ is a function of $\zd$, $\zs$ and $\sigma$. Two images with magnification $\mu_{1,2} = 1 \pm \dfrac{1}{(u/\rein)}$ are created, where $u$ is the size of $\boldsymbol{\vec{u}}$. Thus, Eq. \ref{eq:prob} can be expressed as
\begin{equation}
\begin{aligned}
\fq= &\int_{\zs=\zd}^\infty F(\zs) \: \times \\
&\int_{u=0}^{\rein(\zd,\:\zs,\:\sigma)} n_{\rm s}(\zs, \: L_{\rm lim}/|\mu_2(u)|) \: 2\pi u \: \dd u \: \dd \zs, \label{eq:prob2}\\
\end{aligned}
\end{equation}
which is a simple double integral over $\zs$ and $u$, of a function of $\zd$, $\sigma$, and $L_{\rm lim}$.

To summarize, $\fq$ can be calculated for a given deflector with $\zd$ and $\sigma$, when searching in an imaging survey with limiting magnitude $m_{\rm lim}$. The only prior information required is the redshift dependence of the source LFs, which is used to calculate the number density of sources brighter than $L_{\rm lim}/|\mu_2|$ at angular displacement $\boldsymbol{\vec{u}}$ from the deflector, which is integrated over $\boldsymbol{\vec{u}}$ and $\zs$.

From $\fq$, it is possible to calculate the surface number density of QSO lenses, $\nql$, i.e., the number of QSO lenses in a given area in the sky. We simply add up $\fq$ for all QSO host galaxies within the sky area to obtain $\nql$. Because $\fq$ is a function of $\zd$ and $\sigma$, and assuming that all QSOs are homogeneously distributed, VDFs of QSO hosts can be used to simplify the task. Expressing this in equation form,
\begin{equation}
\begin{aligned}
\nql=\int_{\zd} \: \int_{\sigma} \: \fq(\zd,\sigma^\prime) \: \Phi_{\sigma} (\zd;\sigma^\prime) \: \dd\sigma^\prime \: \dd V_d(\zd), \label{eq:nql} \\
\end{aligned}
\end{equation}
where $\Phi_{\sigma} (\zd;\sigma^\prime)$ is the VDF at $\zd$ with velocity dispersion $\sigma^\prime$ and $\dd V_d$ is the differential comoving volume for a unit area in the sky at $\zd$. Thus, $\nql$ can be calculated for a survey depth $\mlim$ as long as the source and deflector populations are well defined.

\section{Source and Deflector Populations} \label{sec:data}

As discussed in the previous section, the redshift evolution of the VDF of the deflector population, and that of the source LFs are required to calculate $\nql$. In this section, we illustrate which deflector and source populations are used for our calculations.

\subsection{Source Population} \label{subsec:source}

\subsubsection{QSOs as Sources}

\begin{figure}
\centering
\epsscale{0.8}
\plotone{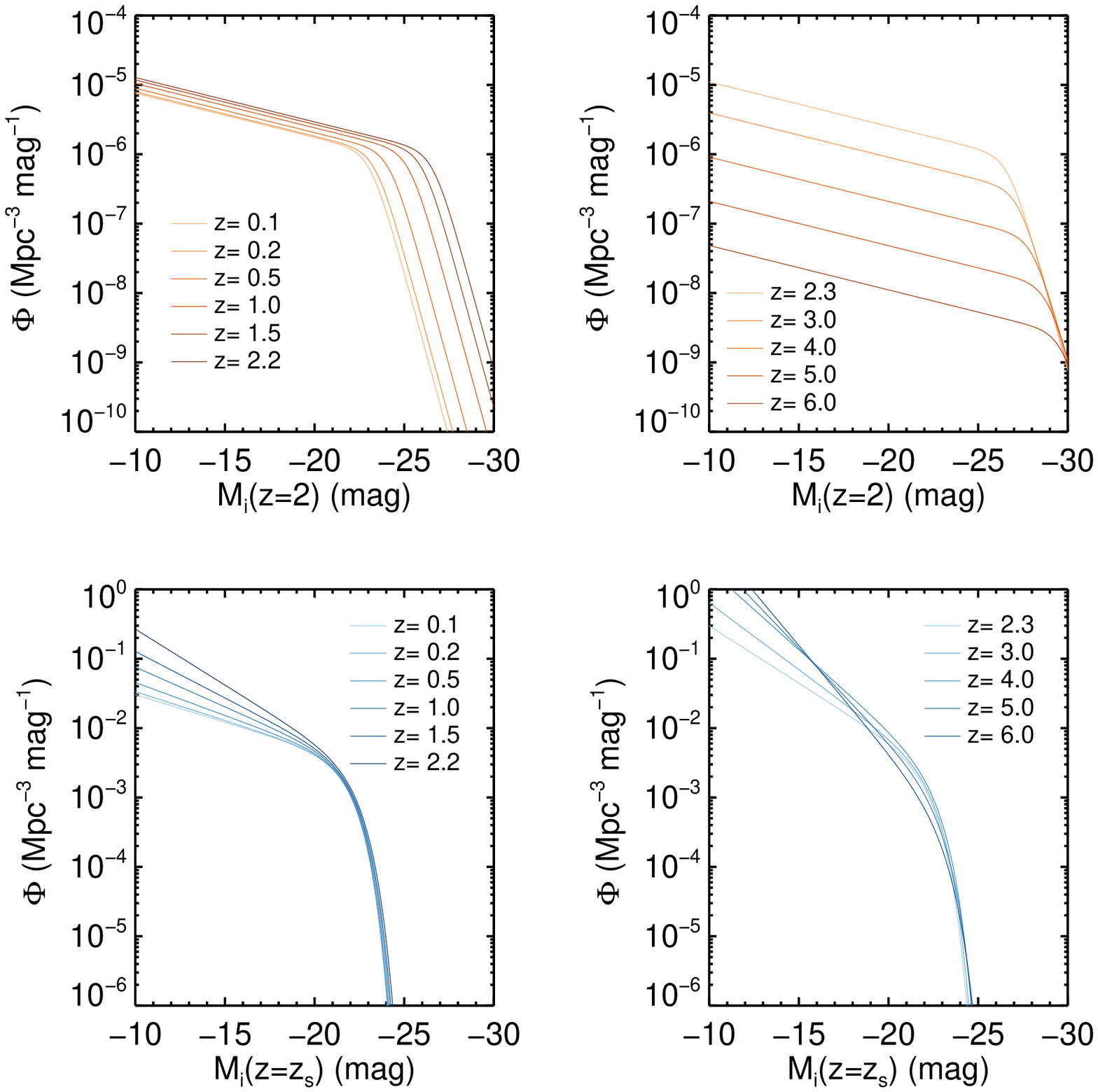}
\caption{QSO (top) and galaxy (bottom) LFs at various redshifts.
}
\label{fig:LF}
\end{figure}

We use the SDSS DR9 QSO LFs from \cite{Ross+13}, who have performed several fits for the redshift evolution of the four parameters ($\Phi^*$, $M_i^*$, $\alpha$, and $\beta$) of QSO LFs in the observed $i$-filter, following a double power-law function of the form 

\begin{equation}
\begin{aligned}
\Phi(M,z)= \dfrac{\Phi^*}{10^{0.4(\alpha+1)(M-M^*)}+10^{0.4(\beta+1)(M-M^*)}}, \label{eq:QLF}\\
\end{aligned}
\end{equation} 
where the redshift dependence is from the four parameters. Among these, we select the parameterizations that fit the data well and are more or less continuous at the redshift 2.2 break; the pure luminosity evolution (PLE) model for $z < 2.2$, and the luminosity evolution and density evolution (LEDE) model for $z > 2.2$ (orange and black lines in Figure 15 of \citealt{Ross+13}). The two fits are slightly offset at the $z=2.2$ break, so we make the fits continuous by fixing the $z=2.2$ parameters to be those from the $z < 2.2$ fit, since this is the fit that is more trustworty, as can be seen from the prediscussed figure of \cite{Ross+13}. This functional form is shown in Table \ref{tbl:QLF}. Although the $z > 2.2$ fit is for up to redshift 3.5 only, which was the limit for the data available at that time, we extend this up to redshift 7, which is where the most distant quasars are being found, while also being roughly the limit of Lyman-break sources that are detectable in the $i$-filter. The actual QSO LFs are shown in the top panels of Figure \ref{fig:LF}.

The magnitudes given here are for the observed $i$-filter for QSOs at $z=2$, which is the standard introduced in \cite{Richards+06}. These magnitudes need to be K-corrected to the observed $i$-filter at the source redshift, which can be done assuming a power-law SED for QSOs, using the prescription described in \cite{Richards+06} as follows:

\begin{equation}
\begin{aligned}
M_i(z=\zs)  &= M_i(z=0) + K_{\rm cont}(\zs) + K_{\rm em}(\zs)\\
&= M_i(z=0) - 2.5(1+\alpha_\nu) \log\: (1+\zs) + K_{\rm em}(\zs)\\
&= M_i(z=2) + 2.5(1+\alpha_\nu) \log\: \Big( \dfrac{1+2}{1+\zs}\Big) + K_{\rm em}(\zs) \\
\end{aligned}
\end{equation} 
where $M_i(z=z')$ is the absolute magnitude observed in the $i$-filter for QSOs at $z=z'$, $\alpha_\nu$ is the spectral slope for $f_\nu$ (i.e., $f_\nu \propto \nu^{\alpha_\nu}$), and $K_{\rm cont}(\zs)$ and $K_{\rm em}(\zs)$ are the K-corrections due to the power-law continuum and QSO emission lines, respectively, as defined in \cite{Richards+06}. The total K-correction ($K_{\rm cont} + K_{\rm em}$) is given in Table 4 of \cite{Richards+06}.

\subsubsection{Galaxies as Sources}

The QSO LFs given in \cite{Ross+13} are optimal for our predictions, in that the LFs are given for a certain \textit{observed} wavelength range (namely the observed $i$-filter for objects at $z=2$); we wish to calculate the expected number of QSO lenses observable in a certain filter. On the contrary, galaxy LFs in the literature are usually given with respect to their \textit{restframe} wavelengths. Thus, we compiled multiple galaxy LFs for various wavelengths at various redshifts, that correspond to the observed $i$-filter wavelengths. For instance, the $V$-filter LF is used for $\zs \sim 0.4$ QSOs, $\sim 1500\AA$ for $\zs \sim 4$, and so on. The list of compiled galaxy LFs is given in Table \ref{tbl:GLF1}. 

We compute the K-corrected absolute magnitudes of the LF break, $M_i^*(z=\zs)$, as follows:

\begin{equation}
\begin{aligned}
M_i(z=\zs) &= M_i(z=0) - 2.5 \log\: (1+\zs) - 2.5 \log\: \Big( \dfrac{f_\nu(\lambda_{i,\zs})}{f_\nu(\lambda_{i,0})} \Big) \\
&= M_F(z=0) - 2.5 \log\: \Big( \dfrac{f_\nu(\lambda_{i,0})}{f_\nu(\lambda_{F,0})} \Big) - 2.5 \log\: (1+\zs) - 2.5 \log\: \Big( \dfrac{f_\nu(\lambda_{i,\zs})}{f_\nu(\lambda_{i,0})} \Big) \\
&\approx M_F(z=0) - 2.5 \log\: (1+\zs) 
\end{aligned}
\end{equation} 
where $M_F(z=z')$ is the absolute magnitude observed in filter F emitted from sources at redshift $z'$, and $\lambda_{F,z'}$ is the central wavelength of filter F redshifted from $z'$ so that $\lambda_{F,z'} = \lambda_{F,0}/(1+z')$.
The final approximation holds because the filter wavelengths are selected to be redshifted to the restframe $i$-filter wavelength, i.e., $\lambda_{i,\zs} \approx \lambda_{F,0}$.

\begin{figure}
\centering
\epsscale{1.0}
\plotone{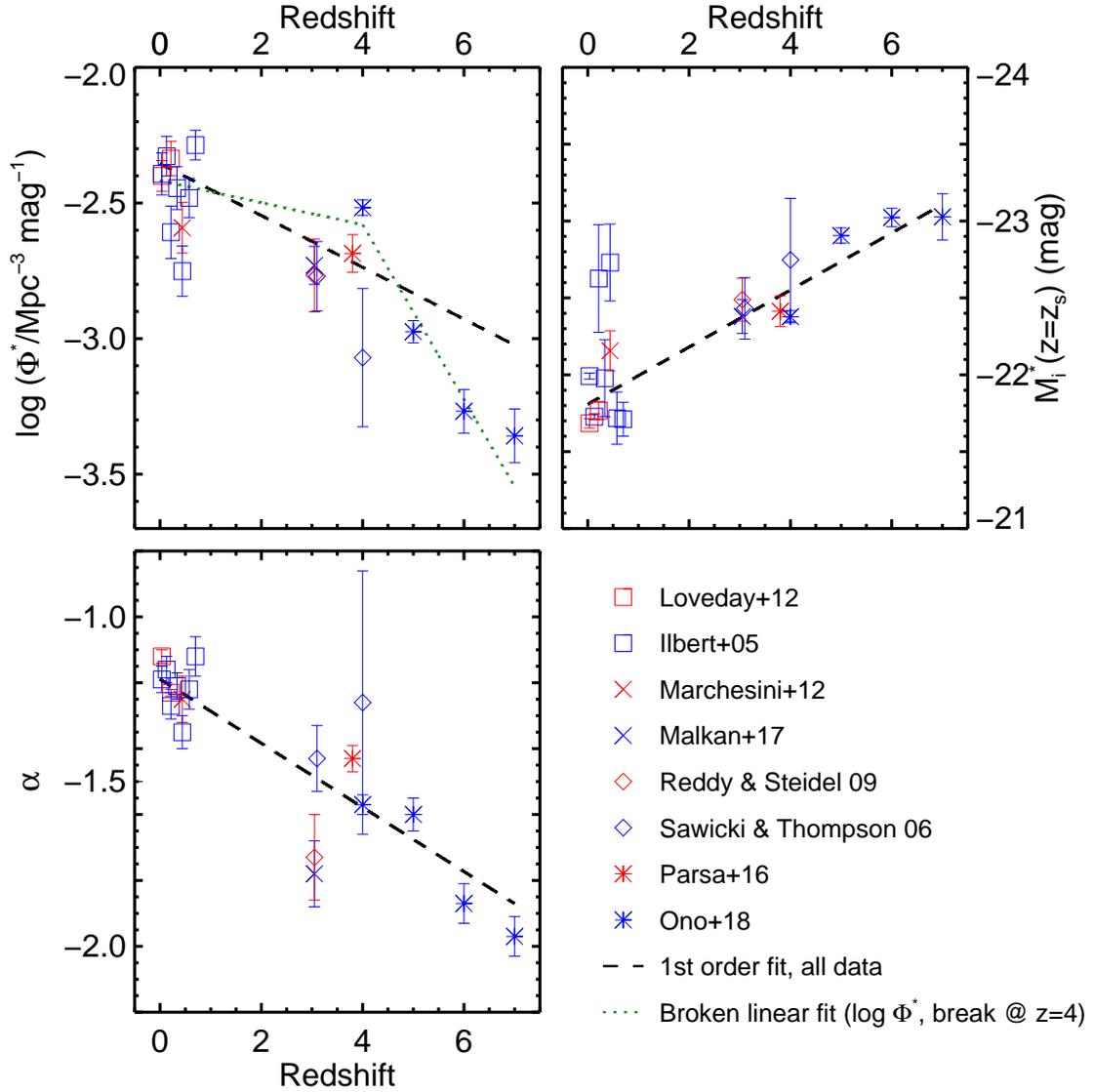}
\caption{Redshift evolution of the three galaxy LF parameters ($\log \Phi^*, M^*, \alpha$), obtained by compiling numerous galaxy LFs for wavelengths at various redshifts that correspond to the observed $i$-filter. Each symbol represents galaxy LFs from separate surveys. The dashed black line indicates the best-fit linear evolution, while the dotted green line is for the two-piece linear fit for $\log \Phi^*$.
}
\label{fig:GLF}
\end{figure}

Since a continuous functional form (with respect to redshift) is desired for computational purposes, we fit the redshift evolution of the three parameters ($\log \:\Phi^*$, $M_i^*(z=\zs)$, and $\alpha$) of the compiled galaxy LFs, assuming the evolution to be linear, which is the simplest case possible. The linear fit does not suit the $\log \:\Phi^*$ data well, so we employ a two-piece linear fit for it. The redshift evolution and parameterizations are shown in Figure \ref{fig:GLF} and Table \ref{tbl:QLF}. We can see in Figure \ref{fig:GLF} that our parameterized fits the data reasonably well, so these choices are justified. The galaxy LFs are shown in the bottom panels of Figure \ref{fig:LF}.

\subsection{Deflector Population} \label{subsec:deflector}
The deflector population in question is constrained to QSO host galaxies. However, although the VDF for normal galaxies in the local universe is relatively well-known (e.g., \citealt{ChoiY+07,SohnJ+17}), that for active galaxies, and moreover its cosmic evolution, is poorly studied. As an alternative, it is possible to infer the VDF from the active black hole mass function (BHMF) using the $\mbh - \sigma_*$ relation. Yet, again our understanding of the redshift evolution of the BHMF is unsatisfactory. 

Thus, two approaches are used to construct the deflector VDFs. Our main approach is to infer the VDF from the QSO LFs using empirical relations, and the second approach is to derive it simply from the observed QSO population; we will call these VDF 1 and VDF 2, respectively.

\begin{figure}
\centering
\epsscale{1.0}
\plotone{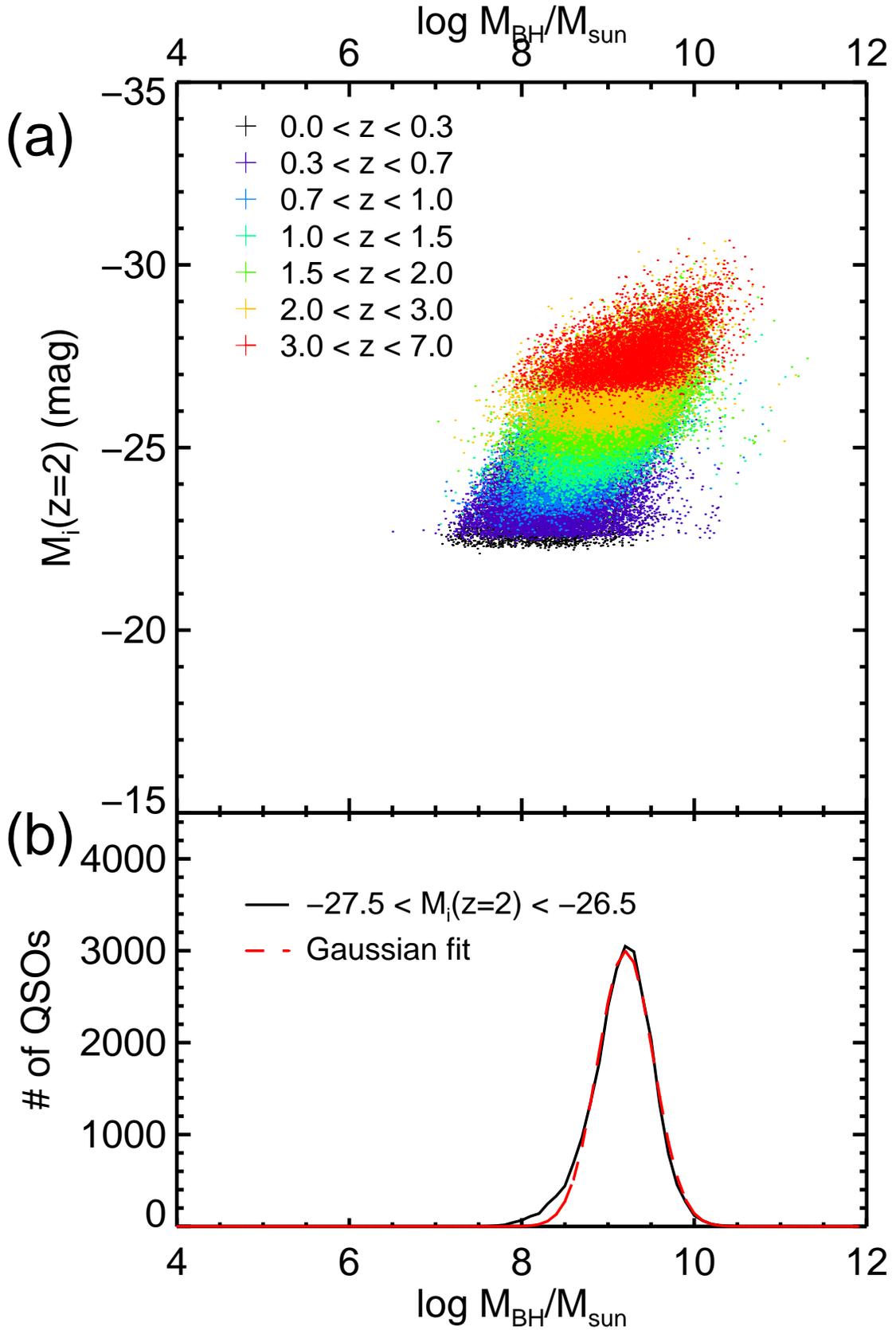}
\caption{(a) $\log\: \mbh$ versus $M_i (z=2)$ for SDSS DR7 QSOs. There is an overall weak yet explicit negative correlation between the two parameters. Each color indicates a different redshift bin. (b) An example of the Gaussian fit to the $\mbh$ distribution for each magnitude bin, shown for the $-27.5 < M_i < -26.5$ bin. We can see that the black histogram is well fit by the red dashed Gaussian function.
}
\label{fig:mbhvsmi1}
\end{figure}

For the first approach, we contemplate that the black hole mass of a QSO can be estimated from its luminosity. Theoretically this is true to some extent; as the single epoch method suggests, more massive SMBHs accrete more mass, and thus they are more luminous. We check this using SDSS DR7 QSOs; Figure \ref{fig:mbhvsmi1}(a) shows the black hole masses from \cite{ShenY+11}, plotted against the $z=2$ absolute magnitude for the $i$-filter ($M_i (z=2)$, hereafter simply $M_i$). We can see that a negative correlation does exist, albeit a weak one. 

To translate the LF into a BHMF, we conduct the following steps. First, the number of QSOs in a specific magnitude bin is obtained from the LF. Then, we fit the $\mbh$ distribution of these QSOs using a Gaussian function, as can be seen in Figure \ref{fig:mbhvsmi1}(b). From this, the fraction of QSOs within a specific $\mbh$ bin among the QSOs in the specified magnitude bin can be estimated by integrating the normalized fitted Gaussian function over the $\mbh$ bin. Finally, we multiply this fraction with the number of QSOs in the magnitude bin, and integrate this over all magnitudes to calculate the number of QSOs in the prespecified $\mbh$ bin. Expressing this in equation form,

\begin{equation}
\begin{aligned}
\Phi_{\mbh} \: \Delta \mu = \int_{-\infty}^\infty \Phi_{M_i} \: \Big[ G(\overline{\mu},\sigma_{\mu}; M_i) \: \Delta \mu \Big] \: \dd M_i ,\\
\end{aligned}
\end{equation}
where $\mu = \log \:(\mbh/M_{\rm sun})$, $\Phi_{\mbh} \: \Delta \mu$ is the BHMF between $\mu$ and $\mu+\Delta\mu$, and $G(\overline{\mu},\sigma_{\mu};M_i)$ is the normalized Gaussian function fit for the magnitude bin at $M_i$, with a mean value of $\overline{\mu}$ and standard deviation of $\sigma_{\mu}$, both of which we can expect to differ for each $M_i$ bin.

\begin{figure}
\centering
\epsscale{1.0}
\plotone{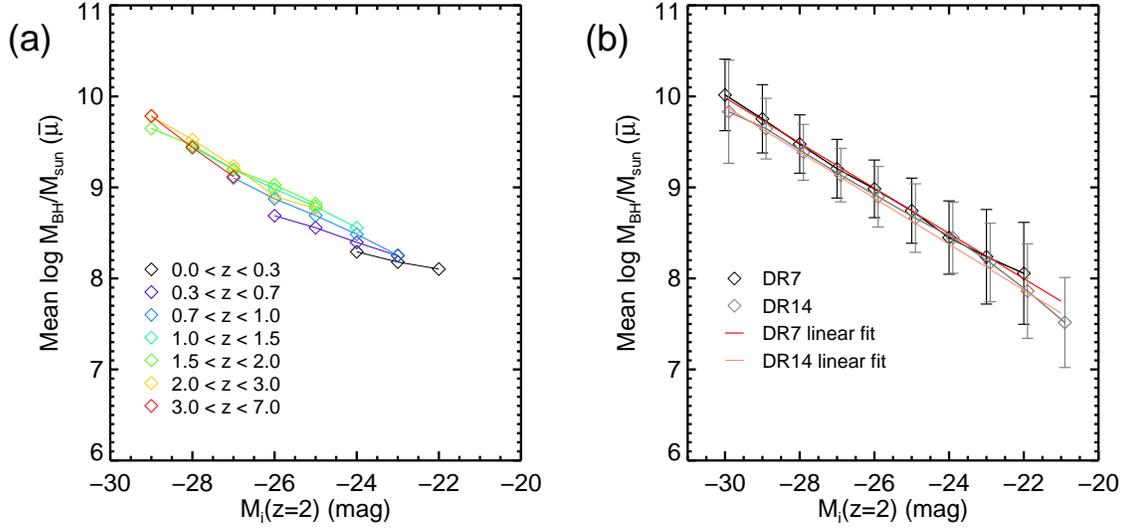}
\caption{(a) Mean values of the best-fit Gaussians of $\log\: (\mbh/\msun)$ $(\overline{\mu})$ for various redshift and absolute magnitude bins. The dependence on redshift is very weak. (b) Mean values of the best-fit Gaussians of $\log \:(\mbh/\msun)$ $(\overline{\mu})$ for various absolute magnitude bins, for all redshifts. Fits for both the SDSS DR7 and DR14 QSO samples are shown. The DR14 data (but not the fit) are shifted by 0.1 in the $x$-direction for clarification.
}
\label{fig:mbhvsmi2}
\end{figure}

We can also postulate that $\overline{\mu}$ depends on redshift. However, Figure \ref{fig:mbhvsmi2}(a) demonstrates that the variation in $\overline{\mu}$ for different redshift bins is relatively small compared to that for different magnitude bins, so a single value of $\overline{\mu}$ is used for all redshifts in a given $M_i$ bin; the result is shown in Figure \ref{fig:mbhvsmi2}(b). In addition, $\sigma_{\mu}$ is roughly similar for all magnitude bins when the QSO sample in the bin is sufficiently large, so this is also fixed to 0.3 dex for future calculations. 

Subsequent to DR7, later generations of SDSS (BOSS/eBOSS; \citealt{Dawson+13,Dawson+16}) have delved into the more distant and less luminous populations of QSOs. Since QSO luminosities are directly linked to the SMBH masses, as shown above, our work, which employs the redshifts and masses of QSO host galaxies, may be significantly affected by these recent changes. Thus, we use the most recent SDSS DR14 QSO catalog \citep{Paris+18} to check whether these recent updates affect our results.
The black hole masses of these QSOs are measured via the single epoch method, and will be published in a subsequent paper (Taak et al. in preparation). Fortunately, Figure \ref{fig:mbhvsmi2}(b) shows that the two QSO samples give similar results in all magnitude bins. Since the DR7 QSOs are the more complete sample, we choose to use the DR7 fit.

The same method is used to translate the BHMF to the VDF of the QSO host galaxies, using the $\mbh - \sigma_*$ relation from \cite{Kormendy+13}, in the form of

\begin{equation}
\begin{aligned}
\Phi_{\sigma} \: \Delta \omega = \int_{-\infty}^\infty \Phi_{\mbh} \: \Big[ G(\overline{\omega},\sigma_{\omega}) \: \Delta \omega \Big] \: \dd \mu, \label{eq:bhtosig}\\
\end{aligned}
\end{equation}
where $\omega = \log (\sigma$/km s$^{-1}$).
$\overline{\omega}$, the mean of the normalized Gaussian function, is obtained from a one-to-one transformation of the $\mbh$ bin using the $\mbh - \sigma_*$ relation. As for $\sigma_{\omega}$, the standard deviation of the Gaussian, the scatter of the relation is used: 0.28 dex in the $\mbh$ direction, or 0.063 dex in the $\sigma_*$ direction.

\begin{figure}
\centering
\epsscale{1.0}
\plotone{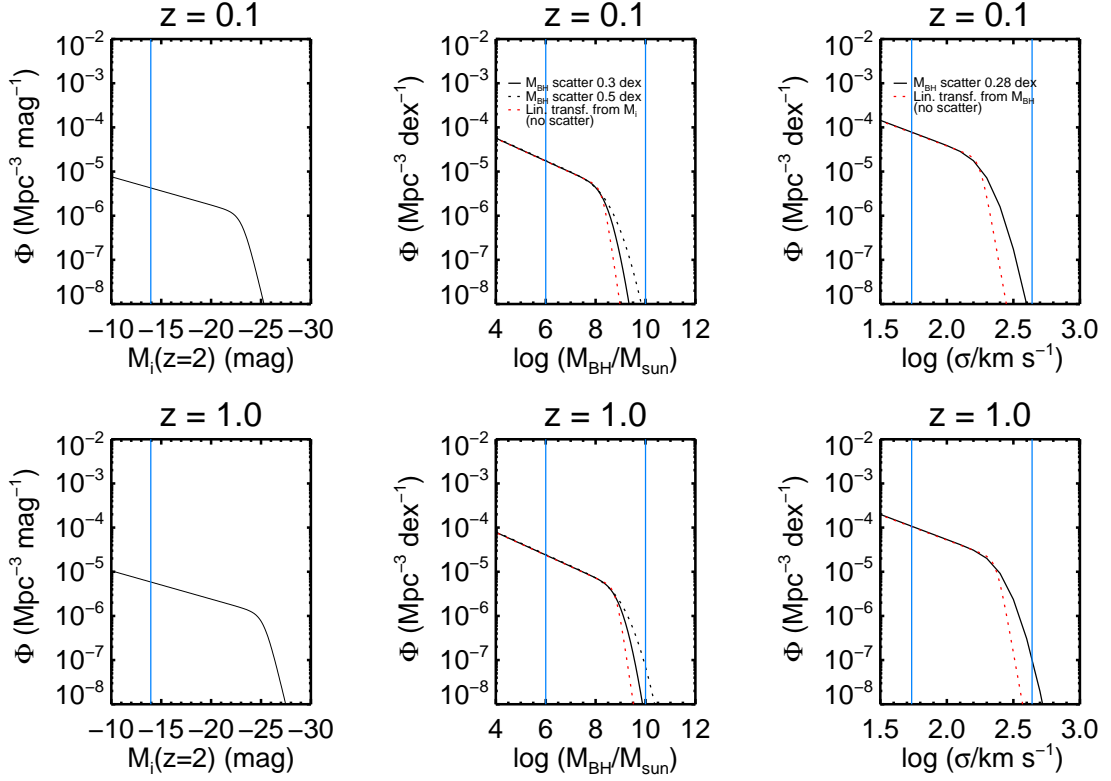}
\caption{QSO LFs (left column), BHMFs (center column), and host galaxy VDFs (right column) for two redshifts, $z=0.1$ and $z=1$. Red dashed lines show the linear translation between the functions, and black solid and dashed lines are for the functions when scatters in the relations are considered. Blue vertical lines show the range for $6 < \log \: (\mbh/\msun) < 10$ in the center column, and the corresponding ranges when using linear translations in the other two columns.
}
\label{fig:qlf_bhmf_vdf}
\end{figure}

Figure \ref{fig:qlf_bhmf_vdf} shows the BHMF derived from the QSO LF of \cite{Ross+13}, and the VDF in turn from the BHMF, for various redshifts. For the transformation from the QSO LF to the BHMF, since the scatter of the correlation ($\sigma_\mu$) may be uncertain, two different scatters are used. As can be expected, the use of a larger scatter yields a larger BHMF in the high-mass bins, but the low-mass bins are not affected by much.

\begin{figure}
\centering
\epsscale{1.0}
\plotone{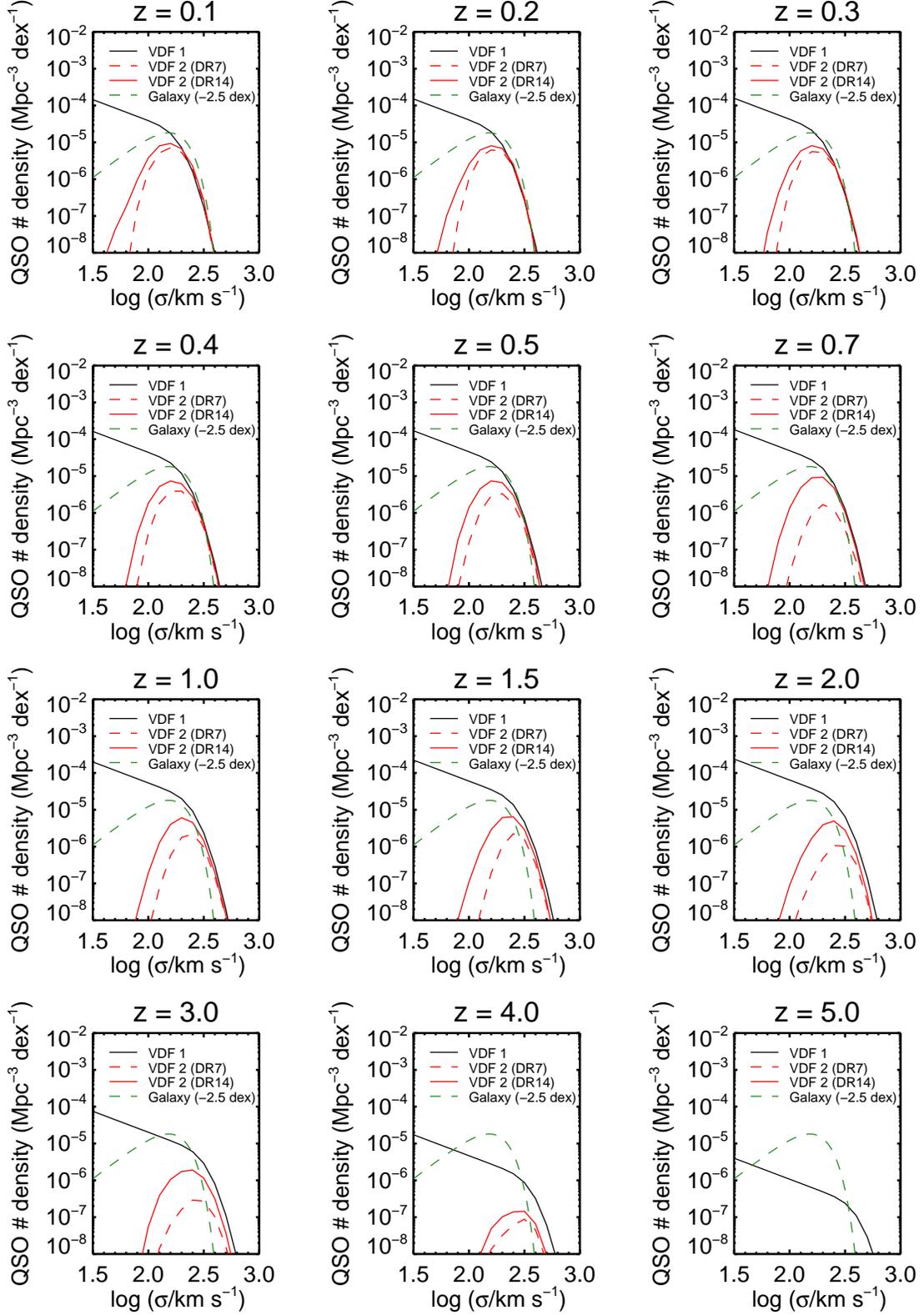}
\caption{QSO host galaxy VDFs at various redshifts. Black lines indicate the VDF estimated from the LF (VDF 1), and red lines show the VDFs from the confirmed QSO sample (VDF 2), with the dashed line for the SDSS DR7 sample and solid line for DR14. The dashed green line is for the galaxy VDF from \cite{ChoiY+07}, decreased by 2.5 dex for comparison.
}
\label{fig:vdfcomp}
\end{figure}

Since VDF 1 is derived from a QSO sample corrected for completeness, we compare these results with VDF 2, using the individual SDSS DR7 and DR14 QSO black hole masses translated to velocity dispersions, with the scatter of the relation applied, in the form of Eq. \ref{eq:bhtosig}. No completeness corrections were made for VDF 2. This is shown in Figure \ref{fig:vdfcomp}, which demonstrates that most of the QSOs residing in high-$\sigma_*$ hosts are observed by SDSS, although the more numerous QSOs in low-$\sigma_*$ hosts are not. 

Thus, following the above steps, the VDF of QSO host galaxies as a function of redshift is estimated sequentially from the QSO LF. This is used, along with the source population parameters, to calculate $\fq$.

\section{Results} \label{sec:results}

In this section, we present the statistics of the probabilities of a QSO host galaxy to be a QSO lens ($\fq$), calculated using the method and data shown in Sections \ref{sec:method} and \ref{sec:data}. 
How $\nql$, the surface number density of QSO lenses, depends on $\mlim$ of the surveys in question is shown. We also derive the probability distribution functions (PDFs) of $\fq$ depending on parameters of the deflector and source populations, such as $\zd$, $\zs$, and $\mbh$. Finally, the number of expected QSO lenses for several surveys are computed.

\subsection{$\fq$, $\nql$ and Probability Distribution Functions} \label{subsec:fq}

\begin{figure}
\centering
\epsscale{1.0}
\plotone{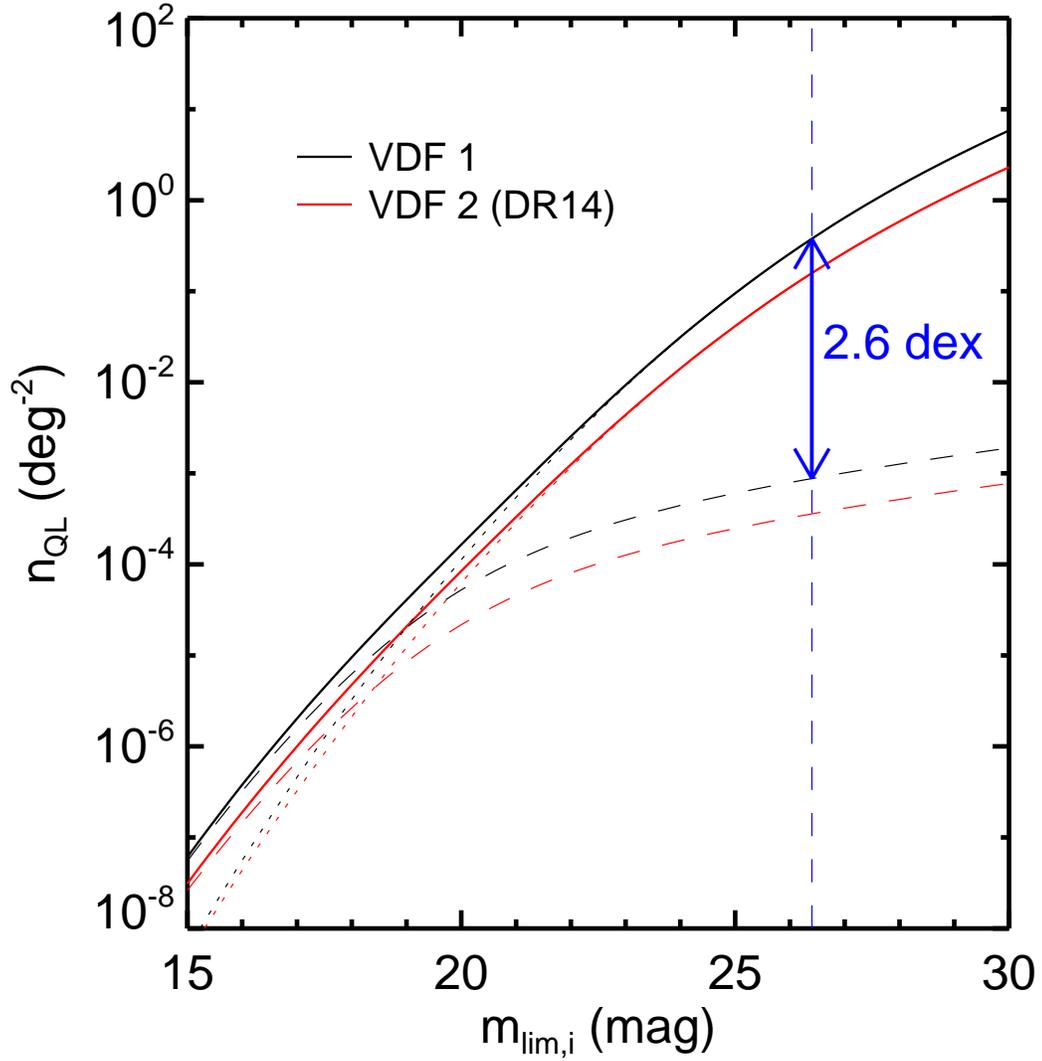}
\caption{$\nql$ versus $\mlim$. Black and red lines indicate whether VDF 1 or VDF 2 (DR14) was used, respectively, and the dotted and dashed lines represent the two source populations, galaxies and QSOs, respectively. The vertical blue dashed line indicates $\mlim = 26.4$ mag, which is the limiting magnitude for the $i$-filter of HSC/Wide DR1, and the double-ended arrow shows the difference in $\nql$ for the two source populations for VDF 1.  
}
\label{fig:fq1}
\end{figure}

Figure \ref{fig:fq1} shows $\nql$ versus limiting magnitude $\mlim$. First of all, as expected, $\nql$ increases for deeper limiting magnitudes. A straightforward explanation is that deeper surveys are able to detect fainter images, which originate from fainter sources, so the number of sources magnified to above the limiting flux increases, and so does the number of lens systems. Second, the two source populations cross at about $\mlim \approx 15.5$ mag. This simply means that below this $\mlim$, QSO sources are more dominant, whereas at the faint end, there are more galaxies than QSOs that will be lensed. Finally, for a limiting magnitude of 26.4 mag, which is the depth of the first data release of the HSC/Wide survey in the $i$-filter, there is a $\sim$ 2.6 dex difference between galaxy and QSO sources. This offset will come up again in the following plots. It should be noted that the lower limits of the limiting magnitudes shown here ($\mlim \sim $ 15--20) are shallower than that for the SDSS QSO sample ($\mlim \approx$ 21 mag), so for these limiting magnitudes, the QSO deflector sample itself will not be fully detected, thus decreasing $\nql$ further. Yet, concurrent surveys have surpassed this limiting magnitude range, and this will be relatively unimportant for the surveys that we discuss here.

\begin{figure}
\centering
\epsscale{1.0}
\plotone{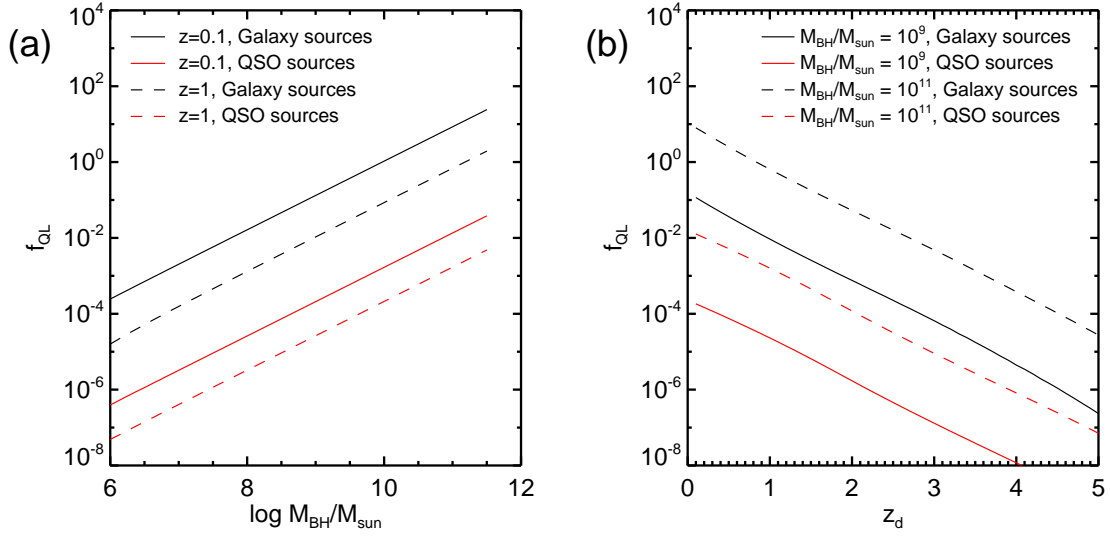}
\caption{$\fq$ as functions of $\mbh$ (a) and $\zd$ (b), for the $i$-filter depth of HSC/Wide DR1 ($\mlim$ = 26.4 mag). Line colors indicate the type of source population, and line styles indicate different deflector redshifts (a) or black hole masses (b).
}
\label{fig:fq2}
\end{figure}

Figure \ref{fig:fq2} shows $\fq$ with respect to parameters of the deflector QSO host galaxies, for $\mlim$ = 26.4 mag. The left plot is $\fq$ versus $\mbh$; we can see that there is a positive correlation between the two. It is possible to infer what the slope of this correlation should be; first, although the actual relation is complicated by the dependence on $\boldsymbol{\vec{u}}$, we can say that $\fq$ is roughly proportional to the ``cross-section'' for sources to be lensed, or $\rein^2$. Also, $\rein \sim \sigma_*^2$, though this is again not straightforward due to $\zs$. Finally, from the $\mbh - \sigma_*$ relation, $\sigma_* \sim \mbh^{0.23}$, so $\fq \sim \mbh^{0.90}$, which is quite similar to the measured slopes of 0.92--0.95 for all four cases shown. In addition, there is a constant $\sim 2.6$ dex offset between galaxy and QSO sources, as was previously shown for Figure \ref{fig:fq1}. This demonstrates that $\mbh$ does not play a critical role in determining this offset. The right panel shows $\fq$ as a function of $\zd$; there is a negative correlation. Simply put, for more distant deflectors, the Einstein volume is smaller, and there are fewer sources that can be lensed. There is also the same $\sim 2.6$ dex offset in this diagram, so this offset is almost entirely due to the differences in the galaxy and QSO source population densities, not because of deflector properties. The offset is not completely constant; this will be discussed in the next paragraph in detail. 

\begin{figure}
\centering
\epsscale{1.0}
\plotone{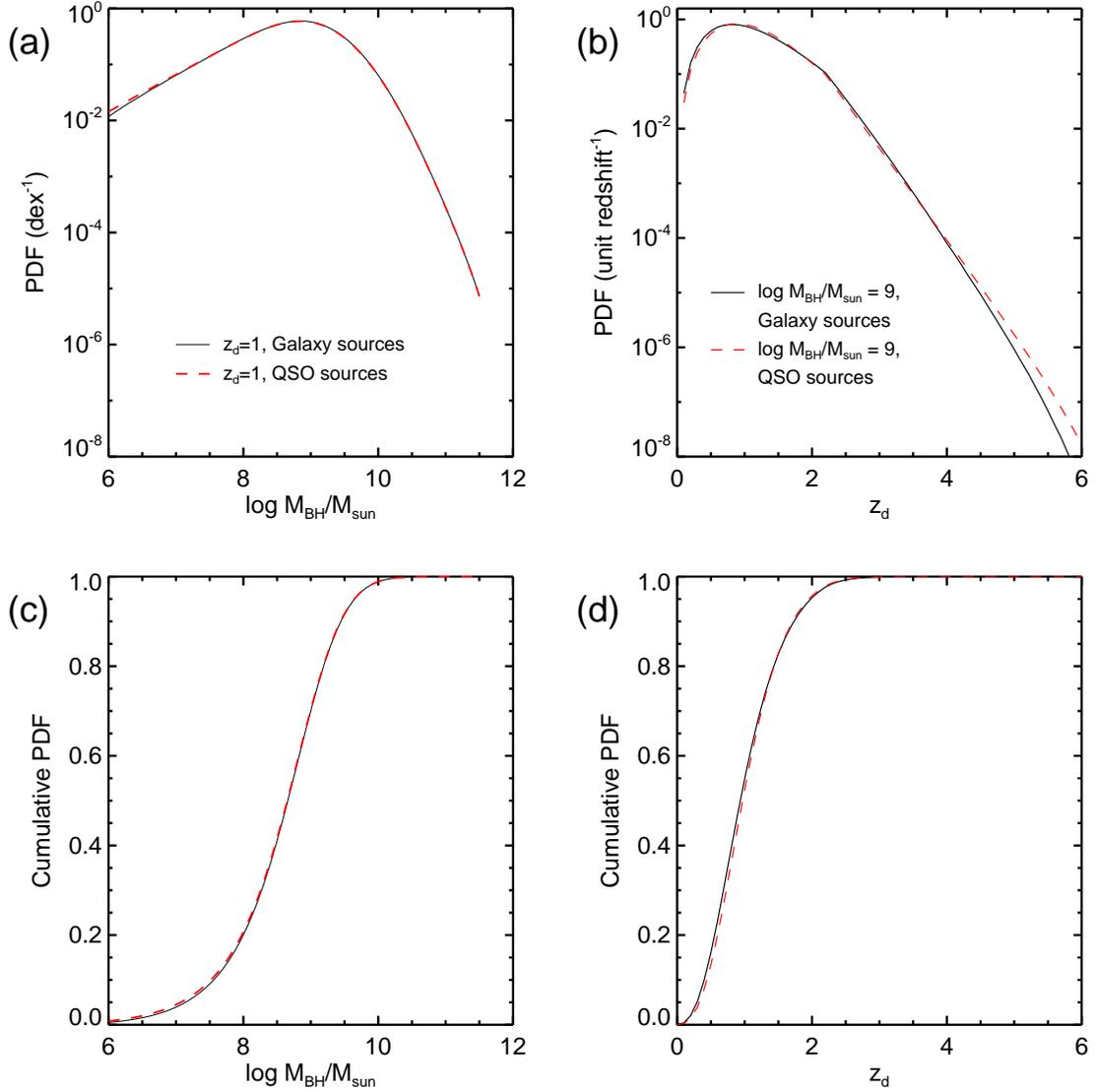}
\caption{PDFs for $\mbh$ (a) and $\zd$ (b), for $\mlim$ = 26.4 mag, the limiting magnitude of the HSC/Wide DR1 in the $i$-filter, and their cumulative distributions ((c),(d)). Black solid lines are for galaxy sources, while red dashed lines are for QSO sources.
}
\label{fig:fq3}
\end{figure}

Figure \ref{fig:fq3} shows the probability distribution of QSO lenses in the top plots, as functions of $\mbh$ and $\zd$, again for $\mlim$ = 26.4 mag, and the cumulative PDF in the bottom plots. From the peaks of the PDFs, we can infer that most of the QSO lenses will be QSOs with $\log \:(\mbh/M_{\rm sun}) \approx 9$ and $\zd \approx 0.7$. It is also interesting to note that about half of all the QSO lenses will have deflector QSOs at $\zd \gtrsim 1$, justifying the main objective of the HULQ project; the furthest QSO lens confirmed so far is at $\zd \approx$ 0.3, and this project will enable us to discover and examine the relatively high-redshift QSO host population. Also, the similarity of each of the two plots in all panels confirm that differences in the source population do not affect the overall PDFs for the deflector parameters. However, if we look at the plots closely, the $\mbh$ panels are almost perfectly identical, while those for $\zd$ are slightly different. This is due to the intrinsic differences in the redshift distribution of the sources; the galaxy number density falls at a steeper rate for higher redshifts when compared to the QSO number density, so the ratio of QSOs that are lensed by high-redshift ($\zd \sim 4$ -- 5) deflectors compared to those lensed by relatively low-redshift ($\zd \sim 1$ -- 2) deflectors is larger than that for galaxy sources.

\subsection{Expected Number of QSO Lenses for Various Surveys} \label{subsec:nql}

As illustrated in the introduction, with sufficiently wide and deep surveys, the present is a good time to start searching for QSO lenses. It is necessary to find out whether this search will be prolific, and which surveys will yield the most QSO lenses to optimize our methods. The calculation of the number of QSO lenses for each individual survey is quite straightforward: $\nql$ for the limiting magnitude of the survey in question is multiplied by its areal coverage. These numbers for several surveys are given in the fourth column of Table \ref{tbl:nqsolens}, along with some survey statistics. The number of detectable QSO lenses is also shown in the final column this table. As will be discussed later in Section \ref{subsec:detect}, these numbers are much smaller than the number of all QSO lenses, due to the image qualities of ground-based surveys.

We can see that among the surveys that have been initiated or completed, HSC/Wide is expected to yield the most QSO lenses with $\sim 440$. Although the numbers in Table \ref{tbl:nqsolens} cannot be simply added up, since some of the surveys overlap with each other, the sample of QSO lenses in currently available imaging data is predicted to be a few dozens, and is expected to reach three digits in the upcoming several years, even with the exception of LSST. 

Since QSO lenses are a deflector-selected sample, discoveries inevitably require searching for lensed features around \textit{confirmed} QSOs. Most of the confirmed QSOs are from the SDSS quasar catalogs, and located in the northern hemisphere, which is also the focus of current and ongoing imaging surveys. Fortunately, several spectroscopic surveys, such as the next phase of SDSS (SDSS-V; \citealt{Kollmeier+17}) and the 4-metre Multi-Object Spectroscopic Telescope (4MOST; \citealt{Merloni+19}) surveys plan to confirm large numbers of QSOs in the southern hemisphere. The main LSST survey, which is scheduled to survey mainly the southern hemisphere, will therefore enable the unveiling of a large number of previously unknown QSO lenses. Space missions, however, are not expected to contain new QSO lenses, since their limiting magnitudes are shallower than ground-based surveys. Yet, their importance in terms of the detectability of QSO lenses will be emphasized in Section \ref{subsec:detect}.

As expected, small surveys below $\sim 100$ deg$^2$ provide only a small number of QSO lenses, despite their relatively deeper imaging. The predicted number depends on the depth and area of the survey; deeper limiting magnitudes lead to larger $\nql$, and the number of SDSS QSOs in a survey is proportionate to its area. In fact, it is possible to determine the relative importance of depth and area, from the slope of the total $\nql$ from Figure \ref{fig:sum}. Consider two surveys, one survey having one tenth of the imaging area of the other but reaching deeper, with identical total exposure times. The smaller survey will have a limiting magnitude 1.25 mag deeper, so assuming that the number of QSOs is proportionate to the area of the survey, $\nql$ must increase by a factor of 10 for a 1.25 mag deeper survey. This results in a slope of $(\log \:10)/1.25 = 0.8$ on the $\log \:\nql$ versus $\mlim$ plot, which is achieved only for $\mlim < 16$ mag. Thus, for a given total exposure time, it is more advantageous to increase the imaging area than to probe deeper to maximize the number of QSO lenses. For $\mlim = 25$ mag, which is a typical limiting magnitude for concurrent surveys, the slope becomes 0.46, indicating that for a survey 10 times smaller, the depth must be $(\log \:10)/0.46 = 2.2$ mag deeper to preserve the number of QSO lenses. For example, the difference in the depths of HSC/Wide DR1 and KiDS is 2.2 mag, and the area coverage of KiDS is 15 times that of HSC/Wide DR1. Thus, the yield from KiDS is 1.5 times that of HSC/Wide DR1. This is the reason for the small surveys having such few QSO lenses; most small surveys that are a set along with larger surveys have areas less than one tenth of their counterparts, but do not reach 2.2 mag deeper limiting magnitudes. Consequently, to search for QSO lenses, it is usually more advantageous to examine larger surveys, and depth is the secondary option.

To summarize, the number of QSO lenses is expected to reach several dozens in currently released data, and will be increased by ongoing surveys within the next few years; LSST will boost that number significantly during its execution. Therefore, the number of QSO lenses is expected be increased from a few tens with currently available data, by several orders of magnitude within a decade.

\section{Discussion} \label{sec:discussion}

\subsection{Comparison with Literature: Galaxy Deflectors} \label{subsec:comp1}

We compare our results with previous studies in the literature discussing galaxy deflectors. Although there are no numerical studies that focus on QSO host galaxies acting as deflectors, \cite{Oguri+10} (hereafter \citetalias{Oguri+10}) have presented the anticipated number of QSOs lensed by galaxies, while \cite{Collett15} (hereafter \citetalias{Collett15}) showed the expected number of galaxy-galaxy lens systems, both for various imaging surveys. Here an A-B lens system indicates that the source B is lensed by the deflector A. We show the deflector and source redshift distributions for potential QSO-galaxy and QSO-QSO lens systems in Figure \ref{fig:zdistrib}, and compare these with the analogous distributions from these two papers.

The left panel of Figure \ref{fig:zdistrib} shows the deflector redshift distributions for galaxy and QSO sources, and the right panel displays the source redshift distributions. We can see that the source population does not affect the potential deflector population much, inferred from the similarity of the distributions. The deflectors have a peak slightly below $z = 1$, while the sources are peaked at $z \approx 2$. These characteristics are similar to those of the distributions shown in Figures 5 and 6 of \citetalias{Oguri+10}, and Figure 1 of \citetalias{Collett15}, which are overplotted with dashed lines. Overall, the distributions are similar to those found in the literature, barring a few notable differences; the PDF of the QSO host deflector redshifts peaks at a further redshift compared to those of normal galaxy deflectors from the literature, and the PDF of the QSO source redshift has a pointy peak in contrast to the other source redshift distributions. The former can be explained in part by the redshift distributions of the deflector populations; as can be seen in Figure \ref{fig:om10comparison}(b), which shows the relative number of QSOs to those of galaxies at a certain redshift, the ratio increases monotonically from $z = 0$ to $z \approx 0.7$. This demonstrates that normal galaxies are much more abundant than QSO host galaxies in the local universe when compared with $z \approx 0.7$, so the redshift distribution of QSO host deflectors should shift to a higher redshift than that for normal galaxy deflectors. The latter is simply due to the evolution of the QSO source population. According to the QSO LFs from \cite{Ross+13}, as can be seen in Figure \ref{fig:LF}, the QSO number density increases monotonically up to $z=2.2$, then decreases for higher redshifts, resulting in an unnatural pointed peak at $z=2.2$, which is propagated directly to the $\zs$ PDF.

In addition, we plot the Einstein radius distribution of the potential lens systems in Figure \ref{fig:eindistrib}. The $\rein$ distribution from \citetalias{Collett15}, overplotted here, is largely similar to our results; their PDF peaks at $\rein \approx 0\farcs4$, and falls by a factor of $\sim$ 10 by $\rein \approx 1\farcs5$. Our results give a slightly smaller $\rein$ peak because the deflector population is different; their galaxy VDF drops off for smaller $\sigma_*$, since it is observationally obtained, while our QSO host galaxy VDF (VDF 1) continuously increases. Notice that there is an overall $\sim$ 2.5-dex difference between the two distributions; this will be explained in the following paragraphs.

The lensed rate $\flensed$, which is the ratio of sources lensed by the deflectors to all sources, is discussed in several papers. \cite{Pindor+03} suggest $\flensed$ is around $10^{-3.4}$, and \citetalias{Oguri+10} proclaim that it is independent of $\mlim$, at $10^{-3.5}$. We plot $\flensed$ for our calculations, for both galaxy and QSO sources in Figure \ref{fig:om10comparison}(a), along with the value of $10^{-3.5}$ given by \citetalias{Oguri+10}. Overall, $\flensed$ for QSO sources is roughly constant over our full $\mlim$ range, which is consistent with \citetalias{Oguri+10}, but there is a $\sim$ 2.5-dex difference. 

We conjecture that this 2.5-dex difference is mostly due to the difference in the number of the two deflector samples, QSO hosts versus all galaxies. We compare the numbers of these two samples; to elaborate, the number of galaxies used as deflectors in \citetalias{Oguri+10} is integrated up to the redshift in question, and the ratio of the number of SDSS DR14 QSOs within this redshift to the aforementioned number of galaxies is calculated as a function of redshift. This graph is plotted in Figure \ref{fig:om10comparison}(c); the logarithm of the ratio fluctuates around $-2$--$-2.5$, thus confirming that the ratio of the numbers of the deflector populations is the dominant factor in the offset. This infers that the mean probabilities for galaxies and QSO host galaxies to be deflectors should be roughly similar. 

In addition, we can conduct a sanity check of our calculations of the number of QSO lenses (Section \ref{subsec:nql}) by comparing with the number of galaxy-galaxy lens systems presented in \citetalias{Collett15}. Details of the source population are different for \citetalias{Collett15}; they use a mock source catalog that is complete up to $i \sim 27.5$ mag, in contrast to our use of LFs with no source luminosity bounds, and a cut of $i = 26.8$ mag for lensed images for the LSST survey. However, considering that magnification will boost the source flux by some factor, \citetalias{Collett15} should give comparable results to our calculations for LSST. The number of galaxy-galaxy lens systems in \citetalias{Collett15} is 11 million for the entire sky. Scaling for the area of the LSST survey gives 4.8 million galaxy-galaxy lenses, and decreasing this by 2.5 dex results in $\sim 15000$ QSO lenses for the LSST coverage. This is similar to $\sim 9700$ resulting from our calculations. \citetalias{Oguri+10} also present the number of galaxy-QSO lens systems for several surveys, many of which overlap with surveys in Table \ref{tbl:nqsolens}. For instance, they predict $\sim 8100$ galaxy-QSO lens systems, which is $\sim 2.7$ dex greater than our results of 15 QSO-QSO lens systems. Numbers for other surveys also agree within a factor of 2--3 (with the exception of PS1/MDS, which reached a shallower depth than what was expected 10 years ago, and is different by a factor of $\sim $4). In all, the number of QSO lenses inferred from the number of galaxy-galaxy and galaxy-QSO lens systems are in line with our results, supporting the robustness of our calculations. 

Overall, our results agree well with previous studies. Distributions of various parameters of the QSO lens population are in good agreement with the literature. The lensed rate, $\flensed$, and the expected number of QSO lenses are also consistent with other papers, assuming that QSO hosts and normal galaxies have similar lens statistics. The main reason behind this is that for the low-$\zd$ ($\zd \lesssim 1.5$) and high-$\sigma_*$ ($\sigma_* \gtrsim 200$ km s$^{-1}$, which corresponds to $\log \:(\mbh/\msun) \gtrsim 8.5$) regimes, where most of the QSO lenses lie (see Figure \ref{fig:fq3}), the VDFs for QSO host galaxies are surprisingly similar to those of normal galaxies, apart from a normalization offset, as can be seen in Figure \ref{fig:vdfcomp}.

\subsection{Comparison with Literature: Imaging versus Spectroscopy} \label{subsec:comp2}

We also compare our results with $\nql$ observed with spectroscopy from the literature. As mentioned in the introduction, the sole comparison can be made with the results of \citetalias{Courbin+10} and \citetalias{Meyer+19}. To reiterate, these studies have looked for candidate QSO lenses using SDSS QSO spectra, searching for emission lines that originate from redshifts further than that of the QSO, in the QSO-subtracted residual spectra. The differences between our work and these studies is that i) we are using imaging data, as opposed to spectral data, so the depths of the data will have distinct meanings; ii) imaging allows us to look for lensed images at large angular distances from the deflector QSO, while spectra can only detect those within the fiber aperture; and iii) imaging requires at least two separate lensed images to confirm as a lens system, but spectra accumulates flux from all images lying within the aperture. Therefore, the number of QSO lenses estimated in this work must be adjusted for the fiber size, the depth of the spectra, and the total magnification to compare with the number of spectroscopically observed QSO lenses.

First, we recalculate $\fq$ for the SDSS fiber size. As an initial calculation, we simply alter Eq. \ref{eq:prob2} to consider only lensed images within the radius of the fiber as follows:
\begin{equation}
\begin{aligned}
\fq= &\int_{\zs=\zd}^\infty F(\zs) \: \times\\ 
&\int_{u=0}^{\rein} n_{\rm s}(\zs, \: L_{\rm lim}/|\mu_{\rm tot}(u)|) \: 2\pi u \: \dd u \: \dd \zs, \label{eq:prob3}\\
\end{aligned}
\end{equation}
where $\mu_{\rm tot}$ is the sum of magnifications of all images inside $R_{\rm SDSS}$, which is the SDSS fiber radius, which we simply assume to be $1\arcsec$, the radius of the BOSS fiber, with which most QSO spectra were taken. We have assumed that the fibers are centered on the targeted deflector QSOs, and the lensed images have PSFs in the shape of two-dimensional $\delta$-functions. For comparison, we also calculate the above for fiducial fiber sizes of 0\farcs5 and 0\farcs1 radii. The results are illustrated in Figure \ref{fig:aper}; we can see that larger fibers are able to gather light from lensed images that are further away from the deflector QSO, so they have larger $\fq$, and consequently yield more QSO lenses. It is worth noting that the $0\farcs5$ fiber and imaging produces almost identical results. For an SIE lens, as we have used throughout this paper, the dimmer image is inside the Einstein radius, while the brighter image is outside it. Since the peak of the $\rein$ distribution is at $\sim 0\farcs5$ (Figure \ref{fig:eindistrib}), for the $0\farcs5$ fiber, usually only the dimmer image is inside it. Thus $\mu_{\rm tot}$ becomes just $\mu_2$ for these systems, and Eq. \ref{eq:prob3} becomes identical to Eq. \ref{eq:prob2}. The actual results differ slightly since not all lens systems have $\rein = 0\farcs5$, but the overall results should be quite similar, as is shown here.

The predictions made above are correct if we were to observe with a very large telescope in an atmosphere-free environment, effectively giving very small PSF FWHMs; in reality, atmospheric seeing and telescope optics cause the images to be blurred according to a point-spread function (PSF), and light that belongs to lensed images within the fiber may be pushed outside of the fiber aperture, or vice versa. We take this effect into account, assuming a seeing of 1\farcs5 (the mean seeing for SDSS spectroscopy time; \citealt{Gunn+06}), and use only the light that remains in the fiber after a Gaussian-seeing convolution. The results are shown as dashed lines in Figure \ref{fig:aper}; even for the largest fibers, a non-negligible portion of the image flux is removed, so the number densities drop inevitably by design. Also, the lensed images are not located at the center of the fiber, so this further reduces the light fraction within it. The decrease is more significant when the apertures are small, since for smaller fibers, more light will be expelled from the fiber. 

To compare with the number of QSO lenses from \citetalias{Courbin+10} and \citetalias{Meyer+19}, we need to constrain the deflector sample further. \citetalias{Courbin+10} used the DR7 sample, with an additional constraint of $\zd<0.7$, while \citetalias{Meyer+19} used the DR12 sample, also with the $\zd < 0.7$ constraint; we will call these Samples 1 and 2. We plot the expected $\nql$ as functions of $\mlim$ for these two deflector samples in Figure \ref{fig:apercomparison}(a). We also plot the results for the DR14 QSO catalog with the same redshift cut of $\zd < 0.7$ applied, which we label as Sample 3. Sample 4 is simply the full DR14 sample, shown for comparison purposes, and these results are plotted as well. We can see that $\nql$ for Sample 4 is the greatest, and by a significant margin over the others; this can be mostly attributed to the ratio of the number densities of the deflector QSO samples. We divide $\nql$ by the number density of QSOs for each sample to obtain the average $\fq$ ($\avfq$), to simply compare the properties of the deflectors; these results are shown in Figure \ref{fig:apercomparison}(b). 

\begin{figure}
\centering
\epsscale{1.0}
\plotone{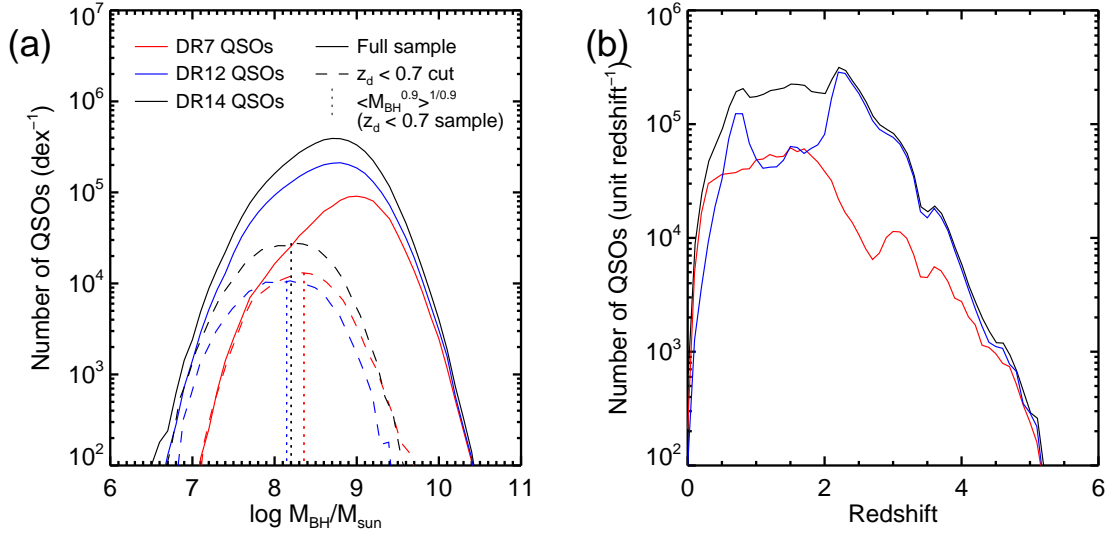}
\caption{$\mbh$ (a) and redshift (b) distributions for SDSS QSOs. Different colors indicate QSO samples from different catalogs. For the left panel, the dashed lines are $\mbh$ distributions for the $\zd < 0.7$ samples, as are used in Section \ref{subsec:comp2}, and the vertical dotted lines indicate the mean values of $\mbh^{0.9}$ for the $\zd < 0.7$ samples, raised to the (1/0.9)-th power for better visualization.
}
\label{fig:mbh}
\end{figure}

\begin{figure}
\centering
\epsscale{1.0}
\plotone{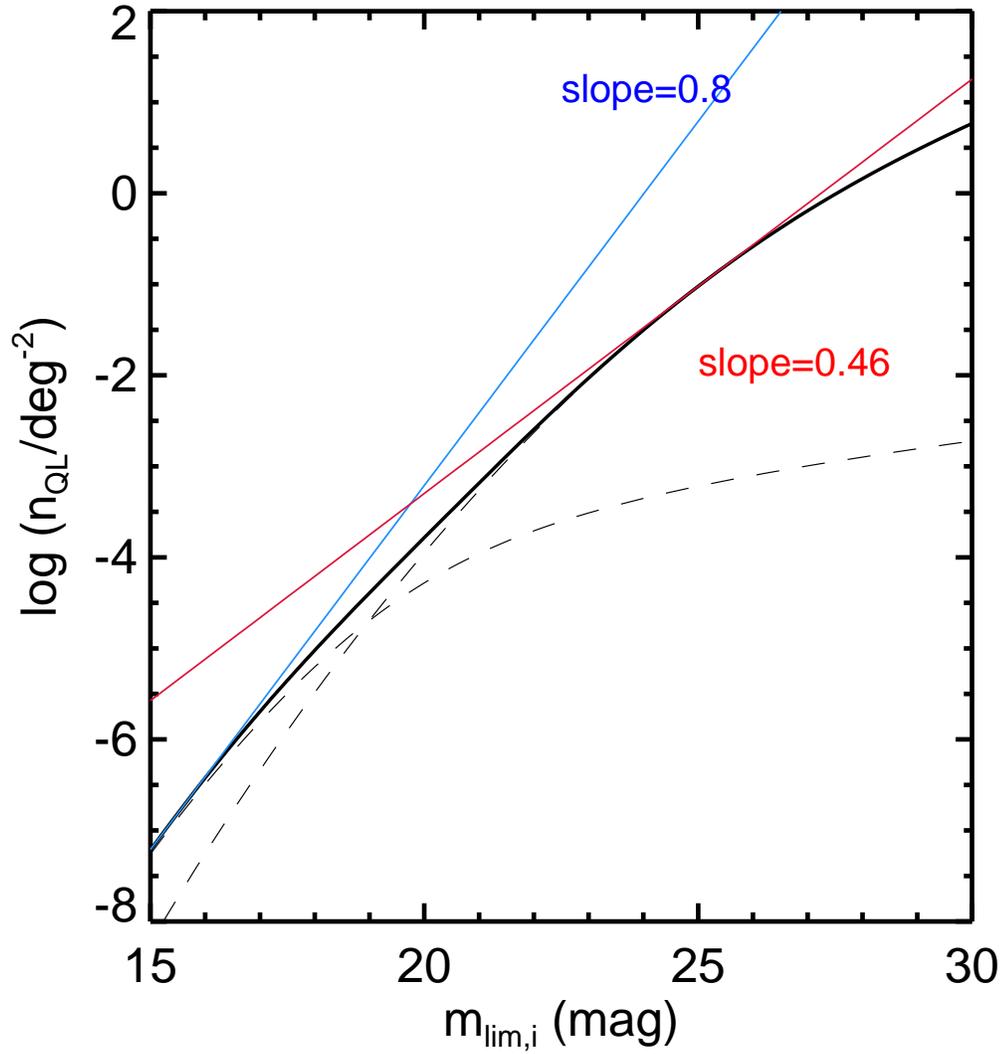}
\caption{$\nql$ as a function of $\mlim$. Black dashed lines show $\nql$ for VDF 1 with galaxy and QSO sources, and the black solid line shows the sum of these two $\nql$, all of which were shown in Figure \ref{fig:fq1}. The blue line indicates where the slope is equal to 0.8, which is only achievable for $\mlim < 16$ mag, and the red line shows the slope of 0.46 at the typical $\mlim$ of recent surveys of $\mlim = $ 25 mag.
}
\label{fig:sum}
\end{figure}

\begin{figure}
\centering
\epsscale{1.0}
\plotone{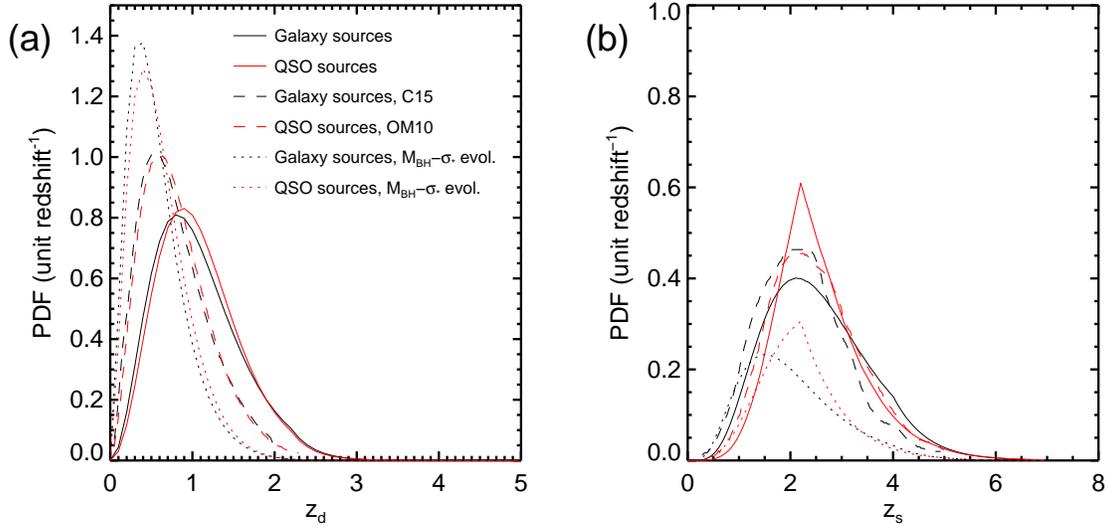}
\caption{The PDFs of the $\zd$ distributions for the deflector QSO host galaxies (a), and those of the $\zs$ distribution for the lensed sources (b), for $\log \: (\mbh/M_{\rm sun}) = 9$ and $\mlim$ = 26.4 mag. The black and red lines are for QSO lenses with galaxies and QSOs as sources, respectively. Solid lines are from this work, dashed lines are from the literature, and dotted lines are for when the evolution of the $\mbh - \sigma_*$ relation is taken into account. The solid lines are identical to the two lines shown in Figure \ref{fig:fq3}(b), but shown in linear scale here.
}
\label{fig:zdistrib}
\end{figure}

\begin{figure}
\centering
\epsscale{1.0}
\plotone{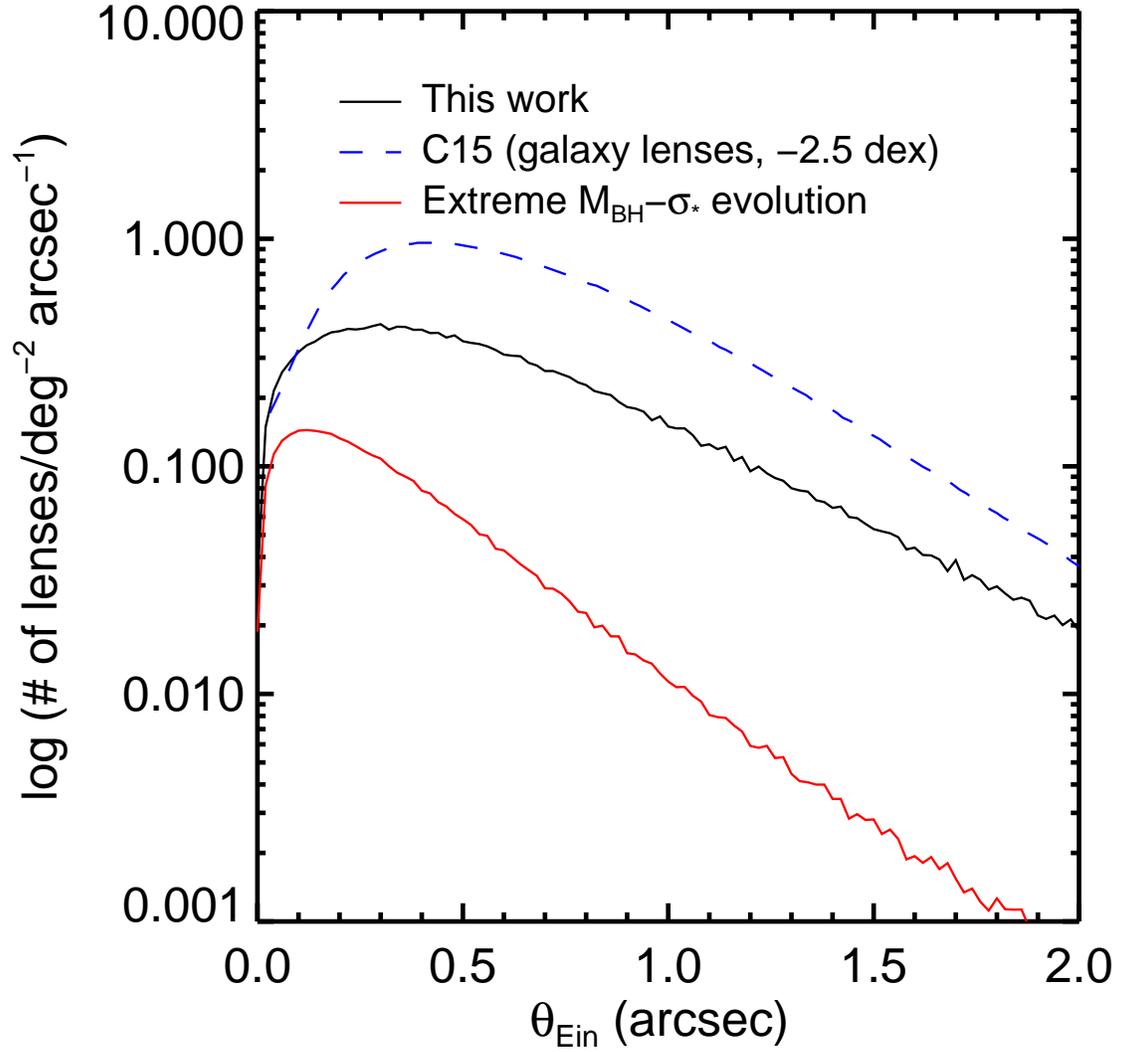}
\caption{The PDF of the $\rein$ distribution for the QSO lenses. The black solid line is from this work, and the blue dashed line is from the literature. The red solid line is for when the evolution of the $\mbh - \sigma_*$ relation is taken into account.
}
\label{fig:eindistrib}
\end{figure}

\begin{figure}
\centering
\epsscale{1.0}
\plotone{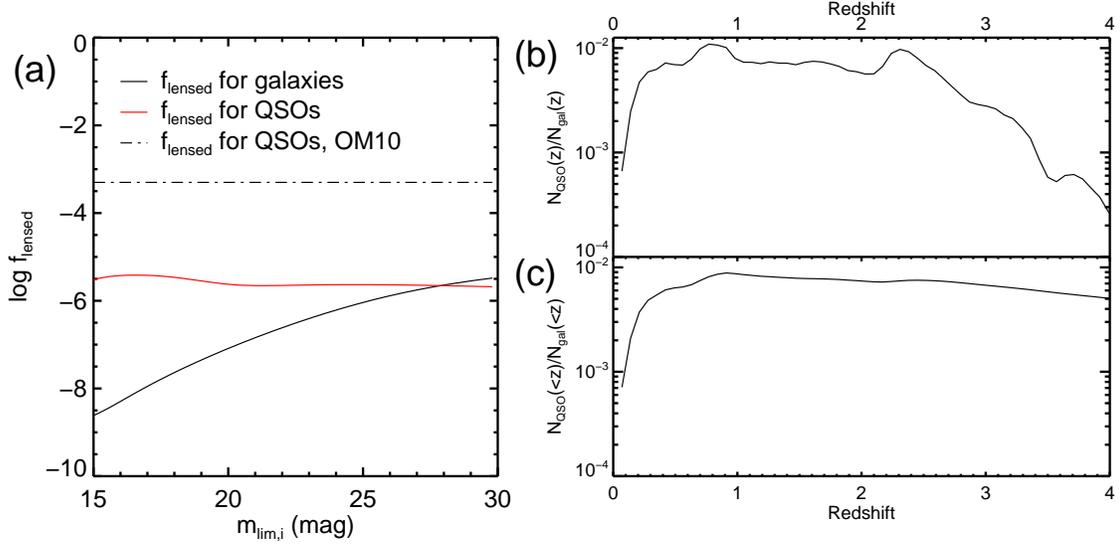}
\caption{(a) $\flensed$ for galaxy (black solid line) and QSO (red solid line) sources, compared with $\flensed$ for QSO sources given by \citetalias{Oguri+10} (black dot-dashed line). The consistency of $\flensed$ independent of $\mlim$ discussed in \citetalias{Oguri+10} is reproduced, but the overall value is $\sim$ 2.5 dex lower. 
(b) The number ratio of QSOs to galaxies as a function of redshift. The number of QSOs at each redshift is directly from the SDSS DR14 QSO catalog, while the number of galaxies is obtained by integrating the VDF of local galaxies from \cite{ChoiY+07} (for $\log \:(\sigma_* $/km s$^{-1}$) $>$ 2), which was used in both \citetalias{Oguri+10} and \citetalias{Collett15}, and then multiplying by the comoving volume at each redshift.
(c) The number ratio of potential deflectors for each \textit{source} redshift, which is the ratio of the cumulative sums shown in (b).
}
\label{fig:om10comparison}
\end{figure}

\begin{figure}
\centering
\epsscale{1.0}
\plotone{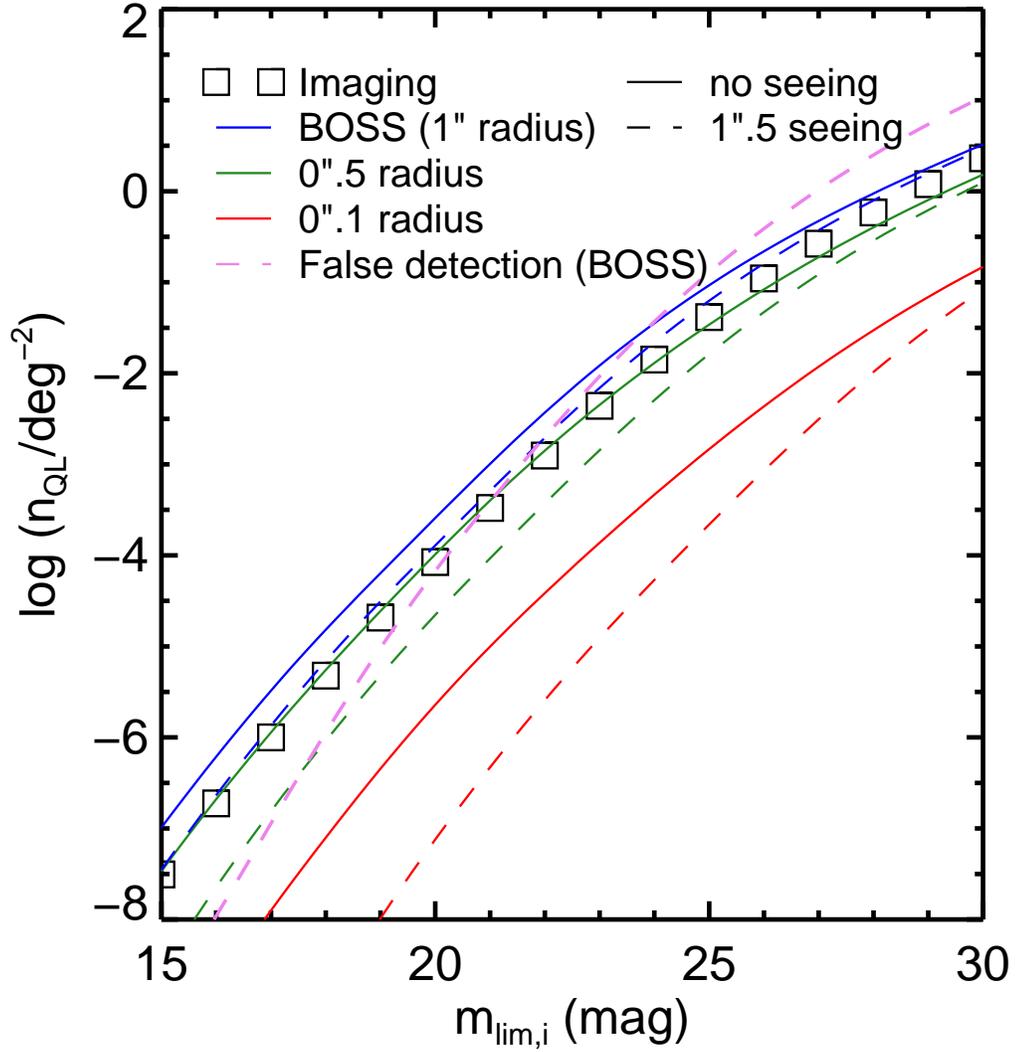}
\caption{$\nql$ versus $\mlim$, for spectroscopy. The black squares show $\nql$ for imaging (for VDF 2 using SDSS DR14 QSOs), and is equivalent to the red solid line in Figure \ref{fig:fq1}. The blue, green, and red lines are for when the size of the fiber is taken into account, for the BOSS fiber (radius of 1$\arcsec$), and fiducial fibers of 0$\farcs$5 and 0$\farcs$1 radii, respectively. Solid and dashed lines are for when seeing is ignored, and a seeing of 1$\farcs$5 (mean seeing at Apache Point Observatory) is used, respectively. The violet dashed line indicates the ``false detection'' number density described in Section \ref{subsec:comp2}, for the BOSS fiber with the seeing applied.
}
\label{fig:aper}
\end{figure}

\begin{figure}
\centering
\epsscale{1.0}
\plotone{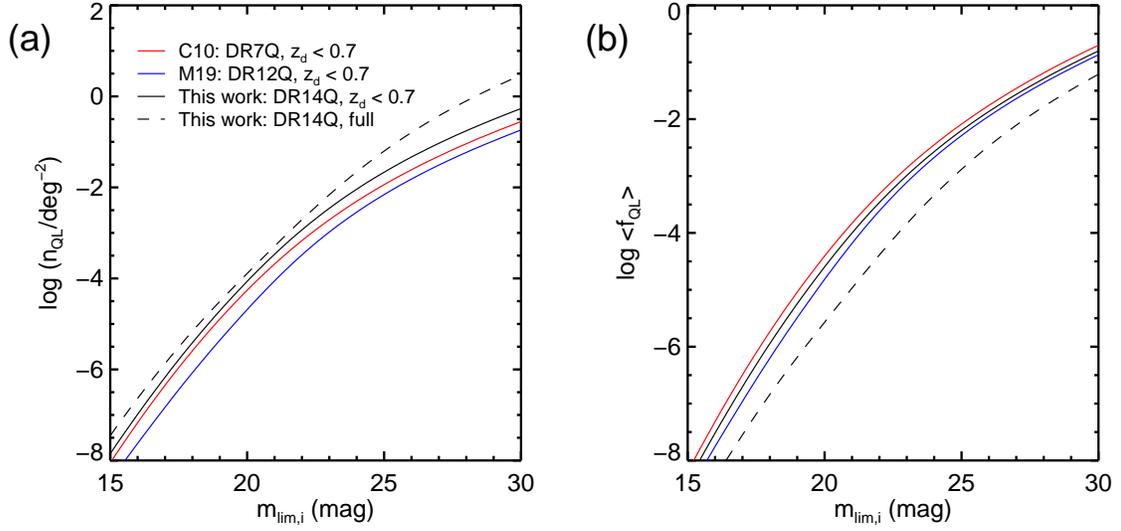}
\caption{$\nql$ (a) and $\avfq$ (b) versus $\mlim$, for some subsamples of the QSO population. The solid red, blue and black lines show $\nql$ for the seeing-corrected BOSS fiber for the three subsamples (Samples 1, 2 and 3, respectively). The dashed black line is for the full QSO sample (Sample 4), plotted for comparison; this line in (a) is equivalent to the blue dashed line in Figure \ref{fig:aper}.
}
\label{fig:apercomparison}
\end{figure}

$\avfq$ decreases in the order of Samples 1, 3, 2 and 4, and this is due to differences in both their $\mbh$ and $\zd$ distributions, which are seen in Figure \ref{fig:mbh}, and illustrated in detail in the following paragraphs. Sample 4 yields the smallest $\avfq$, thus highlighting the importance of the deflector redshifts; as discussed for Figure \ref{fig:fq2}, more distant deflectors have smaller Einstein radii, and have a smaller range for the source redshift (the source must be located \textit{behind} the deflector), so their Einstein volumes are smaller, and so are their $\fq$. A large portion of Sample 4 are located beyond the redshift cut for the other three samples, which leads to the large deviation of $\avfq$ for this sample from those for the rest.

The sample selection for pre-BOSS (SDSS-I/II) QSOs are different from those for BOSS (SDSS-III) and eBOSS (SDSS-IV) QSOs, and the latter surveys have a deeper limiting magnitude, allowing them to probe the low-$\mbh$ regime. Since $\fq \sim \mbh^{0.90}$, we also show the average of $\mbh^{0.90}$ for each sample, to use as a proxy for $\avfq$, in Figure \ref{fig:mbh}(a). We can see that Sample 1 has a larger $\langle\mbh^{0.90}\rangle$ than the other two, explaining why $\avfq$ is larger for Sample 1 compared to Samples 2 and 3.

$\avfq$ for Samples 2 and 3 are somewhat trickier. They have similar $\mbh^{0.90}$, and their redshift ranges are identical. However, a closer look at Figure \ref{fig:mbh}(b) reveals that when compared to Sample 2, Sample 3 has a steeper slope at the low-redshift end ($\zd \lesssim 0.3$), and a more gradual slope at the high-redshift end ($\zd \gtrsim 0.5$) of the sample. This means that the fraction of QSOs at the low-redshift end is greater for Sample 3, and so should their Einstein radii. Consequently, Sample 3 should have a larger $\avfq$ than Sample 2.

Second, the limiting magnitude of the lensed images for the SDSS QSO spectra must be determined. Among the three confirmed lens systems given in \cite{Courbin+12}, we select SDSS J1005+4016, since it is the only system with a lensed image that is deblended relatively easily from the deflector QSO and its host galaxy in \textit{HST} F814W-filter imaging. We first remove the bulge component of the host using the \texttt{ellipse} task of the Image Reduction and Analysis Facility (IRAF). Since the bar of the host passes through the image, we measure the fluxes of two components: the upper bar with the lensed image, and the lower bar. We then take the difference of the two fluxes to obtain the flux of the lensed image, assuming that the bar is symmetric, and then convert this flux to magnitude, which is determined to be 21.1 mag in the F814W-filter. 

\begin{figure}
\centering
\epsscale{1.0}
\plotone{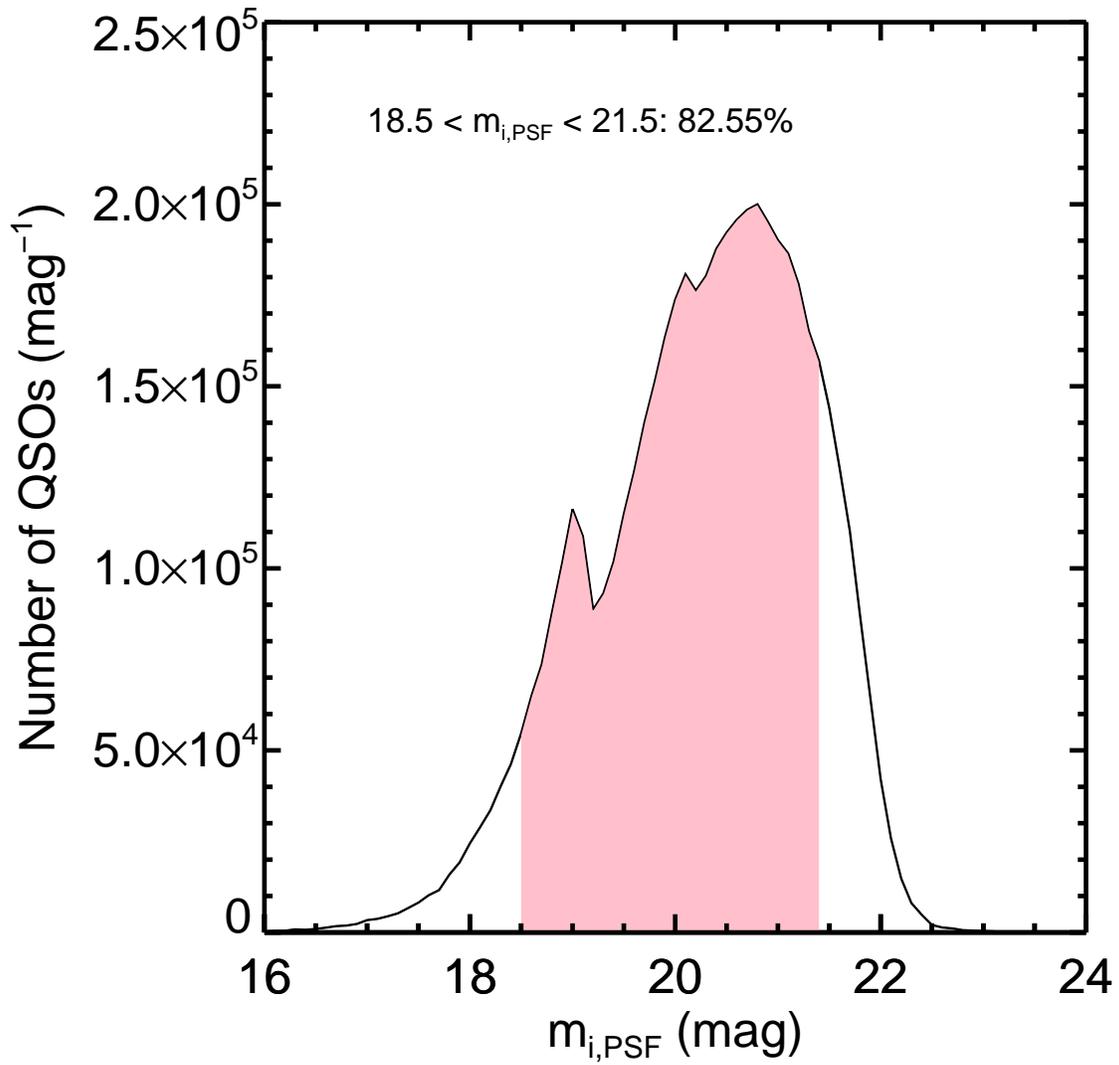}
\caption{$i$-filter PSF magnitude distribution of SDSS DR14 QSOs. Most of the QSOs have $i$ magnitudes between 18.5 and 21.5 mag, shown by the pink fill, and the number of QSOs falls sharply beyond $i \sim$ 21 mag.
}
\label{fig:imag}
\end{figure}

The SDSS spectrum of this system tells us that the spectral lines used to determine this as a QSO lens candidate, [\ion{O}{2}], H$\beta$, and the [\ion{O}{3}] doublet, are close to the detection threshold, with S/N $<$ 6. Thus, we can infer that the ``limiting magnitude'' of SDSS spectra is $\sim 21$ mag in the F814W-filter, and almost identical for the $i$-filter also. Another sanity check can be done with the $i$-magnitude distribution of SDSS QSOs shown in Figure \ref{fig:imag}; it peaks slightly above $\sim 20$ mag and falls off steeply after $\sim 21$ mag, so again this is roughly where the detection threshold is. Although the detection limits for imaging filters and that for discrete emission lines must be different, this value will give a rough estimate of the detection limit of the SDSS spectra.

It is worth noting the case where a source is not multiply imaged, but the single lensed image is within the fiber. These cases will be classified as QSO lens candidates when using spectroscopy, but will not be true QSO lenses. Figure \ref{fig:aper} shows these cases, or ``false detections'', for the BOSS fiber in violet. For the SDSS spectral ``limiting magnitude'' of $\mlim \approx 20$ -- $22$ mag, the number density of false detections are similar to those of actual QSO lenses. This implies that roughly half of the QSO lens candidates discovered via spectroscopy are not \textit{bona fide} QSO lenses.

Finally, we take the value of $\avfq$ from Figure \ref{fig:apercomparison}(b) that corresponds to $\mlim \approx 20$ -- $22$ mag. Sample 1 gives us $-4.4 < \log \avfq < -3.4$, and Sample 2 gives $-4.8 < \log \avfq < -3.7$. The values given from \citetalias{Courbin+10} and \citetalias{Meyer+19} are $-$3.2 and $-$3.5, but these are calculated assuming that all of their candidates (14 and 9) are actual QSO lenses. As discussed above, it is likely that about half of these candidates are non-lenses. In addition, in the case of \citetalias{Courbin+10}, considering that one of the four observed candidates was confirmed as a non-lens, the number of actual QSO lenses ranges from three to 6.5, which gives $-3.9 < \log \avfq < -3.5$ and is more in line with our results for Sample 1. Since no QSO lenses were confirmed in \citetalias{Meyer+19}, their $\log \avfq \approx -3.5$ is simply an upper limit, which is likely decreased by half to $\log \avfq \lesssim -3.8$, so this is also consistent with the Sample 2 results. Thus, our calculations agree with the $\avfq$ values from the spectroscopic searches.

\subsection{Evolution of the $\mbh - \sigma_*$ Relation} \label{subsec:msigmaevol}

There is much controversy about whether the correlations between the properties of SMBHs and their hosts evolve with redshift. It is of general belief that for galaxies with similar masses (or velocity dispersions), the central SMBHs at high redshifts ($z \sim 1$) are heavier than their counterparts in the local universe \citep{PengC+06,Woo+06,Woo+08,Treu+07,DingX+17}. However, whether this trend is sufficiently noteworthy to call it so much as an ``evolution" of such relations is of significant debate \citep{ShenY+15b,Sexton+19}. Since the hosts' velocity dispersions, not masses, are thought to be the main driver for co-evolution (e.g., \citealt{Silk+98}), and the main interest of GL studies is towards velocity dispersions, due to its use when calculating Einstein radii, this section focuses on the possible evolution of the $\mbh - \sigma_*$ relation only.

We adopt the trend from \cite{Woo+08}, whose proposed evolution is the most extreme among various studies, to check how much such a change at higher redshifts affects our results. The redshift dependence of the offset of $\log \:\mbh$ relative to the local relation is as follows:

\begin{equation}
\Delta \log \:\mbh = (3.1 \pm 1.5) \log (1+z) + 0.05 \pm 0.21.\\
\end{equation}

For our calculations, we ignore the $y$-intercept term for agreement with the $z=0$ population. We use this offset from the local relation, whilst ignoring the scatter, and translate it into the \textit{negative} offset of the velocity dispersion, as:

\begin{equation}
\Delta \log \:\sigma_*= \dfrac{1}{4.42} \bigg[-3.1\: \log\:(1+z)\bigg].\\
\end{equation}
We can see that this does not bring a large difference to $\sigma_*$; at $z = 1$, where most of our deflector QSOs are located (Fig. \ref{fig:fq3}b), $\sigma_*$ is decreased by 0.2 dex, and even at $z = 5$, the decrease is about 0.5 dex. Considering that $\fq$ is roughly proportional to the square of $\rein$, which is in turn proportional to the square of $\sigma_*$, a large portion of the deflector QSO hosts will have $\rein$ smaller by $\sim$ 0.4 dex when evolution is taken into account, and thus, we can expect $\avfq$ to be decreased roughly by $\sim$ 0.8 dex.

\begin{figure}
\centering
\epsscale{1.0}
\plotone{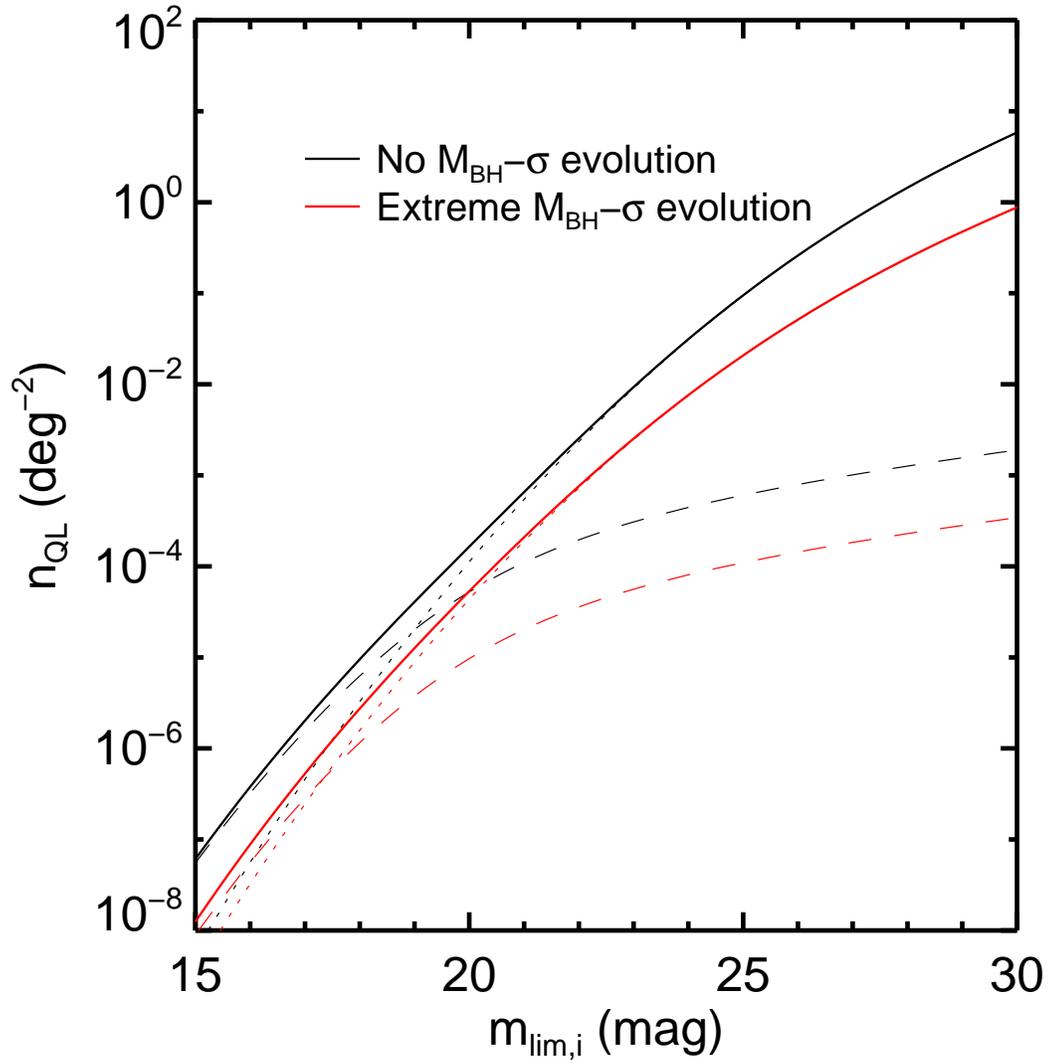}
\caption{$\nql$ versus $\mlim$, when the evolution of the $\mbh - \sigma_*$ relation is considered. Dotted and dashed lines show $\nql$ for the two source populations of galaxies and QSOs, respectively, and the solid line represents the sum of the two. Black lines are for when there is no evolution in the relation, while red lines show the results when the evolution is taken into account. The black lines are identical to those shown in Figure \ref{fig:fq1}.
}
\label{fig:msigma}
\end{figure}

We use this newly-calibrated $\sigma_*$ to calculate a revised set of $\rein$, and in turn, corrected $\avfq$, which is plotted as a function of $\mlim$ in Figure \ref{fig:msigma}. For $\mlim$ = 26.4 mag, $\avfq$ is corrected by a factor of $\sim$ 0.7 dex, which is similar to the above estimate. 
Considering that the evolution of the $\mbh - \sigma_*$ relation that we adopted is the most extreme case, it is reasonable to conclude that the actual correction is much smaller for the surveys we discussed, giving only a minor effect (decrease of a factor of $\sim 2$ -- $3$) on the expected number of QSO lenses, and indicating that the feasibility of the HULQ project is secure.

Now, then, will it be possible for us to infer the evolution of the $\mbh - \sigma_*$ relation from any confirmed QSO lenses from HULQ? If there are a sufficient number of QSO lenses for some redshift range, the $\mbh - \sigma_*$ relation at that redshift can be drawn, with $\mbh$ measurements from single-epoch spectra and $\sigma_*$ from lens analysis. If this is possible for various redshifts, the overall evolution of the relation can be determined.

In the case that the QSO lens sample is plentiful but are distributed rather haphazardly in redshift space, statistics of these lenses can be used to infer the redshift evolution of the relation. First, as was shown in Figure \ref{fig:msigma}, the overall number density of QSO lenses will change depending on the direction and amplitude of the evolution. Also, Figures \ref{fig:zdistrib} and \ref{fig:eindistrib} show the $\zd$ and $\rein$ distributions for when the evolution is taken into account, and compare them with those for the no-evolution scenario. For the $\zd$ distribution, the peak is at a redshift that is significantly lower than that for when the relation is fixed. Similarly, the $\rein$ distribution peaks at $\rein \approx 0\farcs15$, a factor of $\sim$ 2 smaller than that for a constant relation, and decreases faster for larger $\rein$. In addition, the number density of QSO lenses for large values of $\rein$ (i.e., $\rein \gtrsim 1\arcsec$) decreases most significantly (by a factor of $\sim 20$ for this most extreme evolution scenario), so if we can constrain the QSO lens number density accurately, it will be possible to figure out how the $\mbh - \sigma_*$ relation evolves. Finally, if the $\rein$ distribution of the QSO lenses is sufficiently wide, the slope of this distribution offers us another method of checking the evolution of the relation. 

\subsection{Detectability of Lensed Images} \label{subsec:detect}

The calculations executed up to this point assume that all lensed images brighter than the limiting magnitude are detectable. In practice, the lensed images will be close ($\sim 1\arcsec$) to a bright QSO and usually be much dimmer, so they are likely to be buried within the shot noise of the QSO. In order to take this into account, we consider the actual detectability ($f_D$), which depends on a number of parameters, e.g., image quality, brightness of the deflector QSO and lensed images, etc. In general, $f_D$ is a function of five parameters: the deflector QSO magnitude ($m_1$), the lensed image magnitude ($m_2$), the angular distance from the deflector QSO to the lensed image ($\theta_{12}$), the PSF FWHM of the survey, and the limiting magnitude of the survey. For a given survey, the latter two are fixed, so $f_D$ depends on the former three parameters: $m_1$, $m_2$, and $\theta_{12}$.

First, we need to understand how the three parameters affect $f_D$. It is evident that the lensed image is more likely to be detected when the deflector QSO is dim, the lensed image is bright, and the angular distance between them is large. Thus we can expect $f_D$ to be monotonically increasing for $m_1$ and $\theta_{12}$, and monotonically decreasing for $m_2$.

For a survey with a certain depth and a parameter set of $m_1$, $m_2$, and $\theta_{12}$, $f_D$ can be calculated with the following arguments. We construct an image created with the three parameters, and add random noise corresponding to the image quality of the survey, i.e., sky noise, along with Poisson noise from the deflector QSO and lensed images. We fit the image using \texttt{GALFIT} \citep{PengC+02} with a PSF of the survey. Then, we run \texttt{SExtractor} \citep{Bertin+96} on the residual images to check whether the lensed images can be detected. This process repeated for a statistically meaningful number of trials will give us the expected $f_D$ for the parameter set, for a given survey.

The ideal method is to calculate $f_D$ for all possible parameter sets. However, since this is impossible to achieve time-wise, we create an array of the three parameters for which $f_D$ will be estimated, for a single survey; HSC/Wide DR1 in this occasion. Regarding $m_1$, most of the SDSS DR14 QSOs have magnitudes between 18.5 and 21.5 mag (as can be seen in Figure \ref{fig:imag}), so we employ three values of $m_1$: 19, 20 and 21 mag. Since the limiting magnitude of HSC/Wide DR1 is 26.4 mag, the steps used for $m_2$ is from $m_1$ to 26.5 mag, in 0.5 mag steps. Also, $\theta_{12}$ is varied from 0 to $2\arcsec$, with step sizes of one-sixth of a pixel, or 0\farcs028. Each configuration is created 500 times, and $f_D$ is calculated to be the rate of detection among the 500 PSF-subtracted residual images. The lensed image is defined to be ``detected'' when the detected lensed image is less than 0.5 mag fainter than the input lensed image magnitude. A PSF obtained from the HSC PSF Picker is used for the PSF subtraction.

\begin{figure}
\centering
\epsscale{1.0}
\plotone{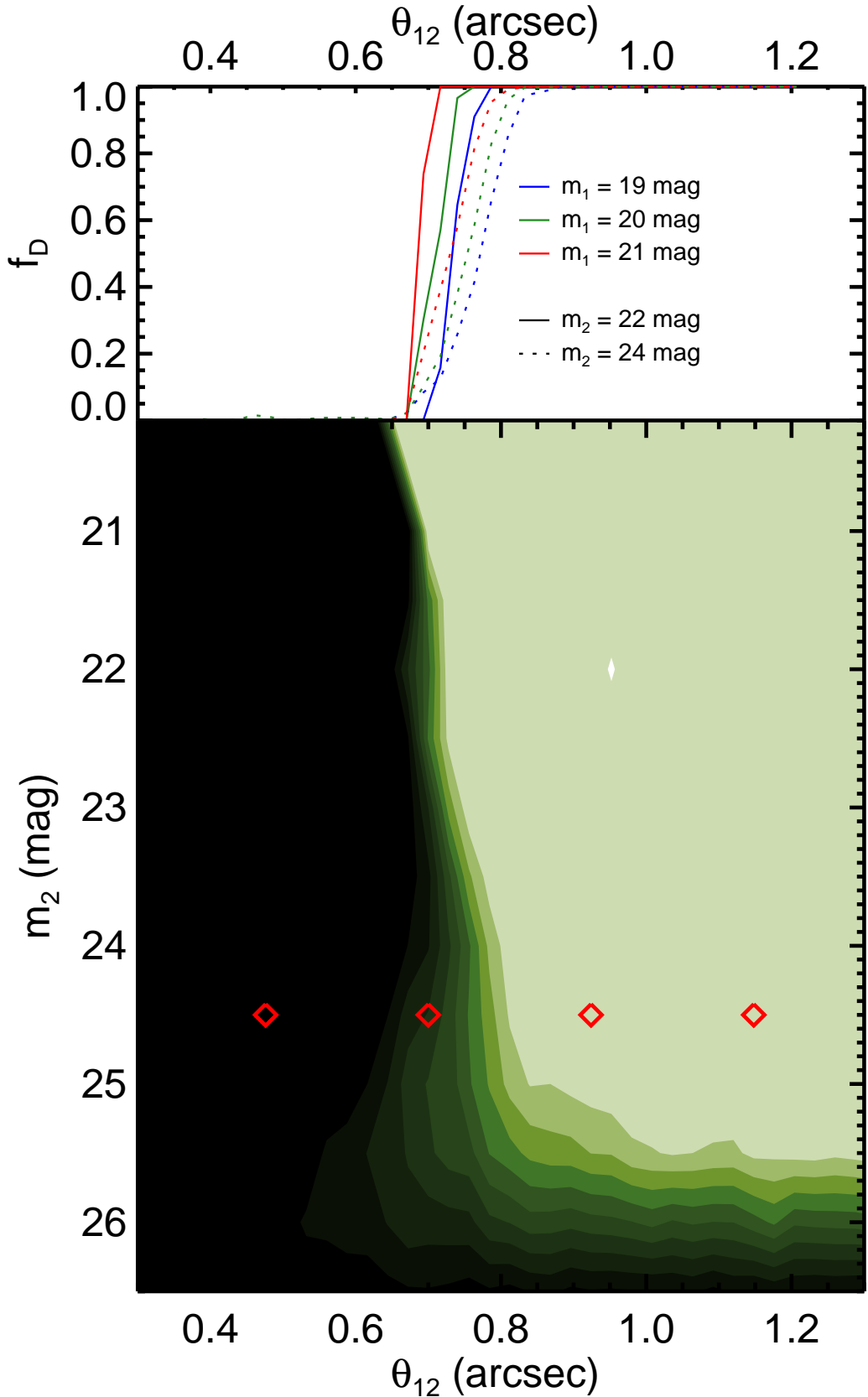}
\caption{(Top) $f_D$ versus $\theta_{12}$, for HSC/Wide DR1. Blue, green and red lines are for $m_1$ (magnitude of deflector QSO) = 19, 20 and 21 mag, respectively. Solid and dotted lines are for $m_2$ (magnitude of lensed image) = 22 and 24 mag, respectively. 
(Bottom) $f_D$ as functions of $m_2$ and $\theta_{12}$ for $m_1$ = 20 mag, for HSC/Wide DR1. Light green represents 100\% detections, whereas black indicates 0\% detections. The red diamonds show the four configurations depicted in Figure \ref{fig:detect2}.
}
\label{fig:detectability}
\end{figure}

This result is shown in Figure \ref{fig:detectability}. The top panel demonstrates that the effect of $m_1$ is relatively insignificant; $f_D$ drops abruptly to 0 for $\theta_{12} \lesssim 0\farcs7$, regardless of $m_1$, for $m_2 = 22$ mag. For dimmer lensed images, the decline is more gradual, and the cutoff moves to slightly larger $\theta_{12}$. Since the dependence on $m_1$ is weak, it is reasonable to assume that the $m_1 = 20$ mag plot, shown on the bottom panel, applies for all QSOs. 

Using the values of $f_D$ calculated as discussed above, we alter Eq. \ref{eq:prob2} to get the modified $\fq$ as 
\begin{equation}
\begin{aligned}
\fq= &\int_{\zs=\zd}^\infty F(\zs) \: \times \\
&\int_{u=0}^{\rein(\zd,\:\zs,\:\sigma)} n_{\rm s}(\zs, \: L_{\rm lim}/|\mu_2(u)|) \: \overline{f_D} \: 2\pi u \: \dd u \: \dd \zs, \label{eq:prob_fd}\\
\end{aligned}
\end{equation}
where $L_{\rm lim}$ is the luminosity corresponding to the limiting magnitude of HSC/Wide DR1, and $\overline{f_D}$ is $f_D$ for the nearest $m_2$ and $\theta_{12}$. For instance, for a deflector QSO of 20.2 mag, and for deflector-source configurations that give a dimmer image magnitude of 24.3 mag and deflector-image angular distance of 1\farcs5, $f_D$ corresponding to $m_2$ = 24.5 mag, and $\theta_{12}$ = 1\farcs512 is used.

We then incorporate this into Eq. \ref{eq:nql} to obtain the number density of detectable QSO lenses, and multiply it by the area of HSC/Wide DR1 to obtain the number of detectable QSO lenses in HSC/Wide DR1 to be 8.0; accounting for $f_D$ has decreased the number of QSO lenses to roughly one-fifth for the image quality of HSC/Wide DR1. It is reasonable to presume that this factor of $\sim 20\%$ should be related to the $\rein$ distribution shown in Figure \ref{fig:eindistrib} and the seeing of the survey; larger seeing values should naturally lead to lower $f_D$ at a given $\theta_{12}$, which is linked to $\rein$. We speculate that the fraction of detectable QSO lenses is equivalent to the fraction of QSO lenses with $\rein$ greater than the seeing multiplied by some factor, and since the fraction of the $\rein$ distribution with $\rein > 0\farcs96$, or 1.7 times the seeing of HSC/Wide DR1, is about 20$\%$, we assume that the fraction of detectable QSO lenses is the fraction of QSO lenses with $\rein$ greater than 1.7 times the seeing value.

Based on this hypothesis, the number of detectable QSO lenses is calculated for each survey and given in the last column of Table \ref{tbl:nqsolens}. The decrease is significant; around ten are predicted to be detectable in currently available imaging data, and before LSST, HSC/Wide is expected to be most fruitful with 82. Still, LSST is complete, the numbers will hopefully reach one thousand, and space missions are forecast to discover a few thousands more. It becomes clear why no QSO lenses have been discovered with ground-based imaging data so far; the most concurrent large survey is PS1/3$\pi$, but its seeing was quite poor, so only $\sim 3\%$ of the hundred or so QSO lens sample in the survey would have been found.

These results also emphasize the importance of space-based surveys for QSO lens discoveries. First, the majority of QSO lenses have $\rein < 0\farcs5$, meaning that surveys with seeings less than $0\farcs3$ are desired to increase the QSO lens sample by factors of several. For instance, CSS-OS is expected to recover $\sim 80\%$ of all QSO lenses with its excellent $R_{\rm EE80} \approx 0\farcs15$, so compared to the $\sim 20\%$ for HSC/Wide DR1, which has the best seeing for ground-based surveys, a much larger fraction of QSO lenses can be discovered by space missions.
Second, these small-$\rein$ systems are necessary if we wish to study the $\mbh - \sigma_*$ relation in detail. As explained in Section \ref{subsec:msigmaevol}, if the $\mbh - \sigma_*$ relation is to evolve with redshift as proposed, most of the QSO lenses will have $\rein \lesssim 0\farcs2$. The position of the peak of the PDF and its slope beyond the peak can be used to constrain the magnitude of the evolution of the $\mbh - \sigma_*$ relation, and the relative number of QSO lenses with small $\rein$ (i.e., the slope of the $\rein$ distribution) will be critical in determining this. 
Finally, this small-$\rein$ subsample allows us to probe the low-mass end of the QSO host galaxy population, thus enabling us to establish the $\mbh - \sigma_*$ relation over a wider range of both $\mbh$ and $\sigma_*$.
To summarize, the upcoming space-based imaging surveys will be a decisive factor in the future of the HULQ project.

\subsection{Sources of Uncertainty} \label{subsec:uncer}

\subsubsection{Scatters of Scaling Relations Used for VDFs} \label{subsubsec:scatter}
In Section \ref{subsec:deflector}, scaling relations were used to translate the QSO LFs to BHMFs, which in turn were transformed to host galaxy VDFs. As can be seen in Figure \ref{fig:qlf_bhmf_vdf}, the size of the scatter of the relations determines the number of QSO host galaxies in the high-$\sigma_*$ end, which are most likely to be QSO lenses. In this section, we demonstrate the importance of these scatters in calculating $\nql$.

Scatters of 0.3 dex and 0.28 dex in the $\mbh$ direction for the $M_i - \mbh$ and $\mbh - \sigma_*$ relations, respectively, were used in Section \ref{subsec:deflector}. We employ scatters of 0.5 dex for both relations for comparison, for which the results are shown in Figure \ref{fig:disc}. As expected, a larger scatter pushes more QSO hosts to higher $\sigma_*$, so more QSO lenses are anticipated. Both increments in the scatters boosts $\nql$ by a similar amount ($\sim$ 0.15 dex at $\mlim$ = 26.4 mag). Therefore, even if the scatters were inaccurately estimated, $\nql$ will not change by a large amount.

\subsubsection{Upper limits to $\sigma_*$}

It is yet uncertain whether galaxies have a limit on velocity dispersions; some studies hypothesize that a hard upper limit to $\sigma_*$ exists ($\sim 400$--$450$ km s$^{-1}$; \citealt{Bernardi+08,Salviander+08}), while others have found galaxies beyond such limits (510 km s$^{-1}$; \citealt{vanDokkum+09}), albeit at high redshifts. Regardless, the number of galaxies with $\sigma_* > 400$ km s$^{-1}$ are rare, and it is possible to presume that such an upper limit exists. In this section, we test how such an upper limit affects our results. 

We modify the VDF so that all QSO hosts with $\sigma_*$ larger than some upper limit have a velocity dispersion corresponding to that upper limit. Similar to the discussion in \ref{subsubsec:scatter}, the high-$\sigma_*$ end is altered, although in the opposite direction, so we can expect $\nql$ to show the opposite behavior as was shown above. Figure \ref{fig:disc} shows that this is indeed the case; $\nql$ decreases when such an upper limit is introduced. However, even an exaggerated cut of $\sigma_* = 300$ km s$^{-1}$ does not result in a significant change in $\nql$ (0.04 dex at $\mlim$ = 26.4 mag), and an impractical upper limit of $\sigma_* = 200$ km s$^{-1}$ is necessary for the reduction to be conspicuous (0.2 dex at $\mlim$ = 26.4 mag). Thus, it is safe to say that upper limits to the QSO host galaxy velocity dispersions do not affect our results at a meaningful level.

\subsubsection{Assumption of Point-Source Lensed Images}
Throughout this paper, we have assumed that all lensed images are point sources, regardless of the morphology of the potential sources. Unfortunately, as Figure \ref{fig:fq1} shows, most of the QSO lenses will have galaxies as sources, which are intrinsically extended, and more so when they become magnified. Therefore, the lensed images will have lower surface brightnesses, and become more difficult to detect; the number of QSO lenses could decrease significantly. In this section, we verify whether this assumption was correct.

For the assumption to hold, the angular size of galaxies must be similar to or smaller than the PSF seeing. The PDF in Figure \ref{fig:fq3}(b) shows that the most common galaxy source redshift is $\zs\approx2$, and the fraction of sources at $\zs < 1$ is meager. The mean galaxy size at $z = 2$ ($z = 1$) is $\sim$ 2 (3) kpc \citep{Ribeiro+16}, which corresponds to an angular size of $0\farcs24$ ($0\farcs37$). So we can reasonably argue that the average galaxy is smaller than the ground-based seeing at $\zs > 1$. Galaxies that are dimmer, which are more important in this argument since they are closer to the detection threshold, must be even smaller \citep{Im+95}. Since these sizes are smaller than the best seeing achievable from the ground, it is safe to say that most galaxy sources can be assumed to be point sources. Obviously, this assumption becomes incorrect when space telescopes are considered, so the numbers given for \textit{Euclid} and \textit{CSST} in Table \ref{tbl:nqsolens} should be thought as upper bounds.

In addition, gravitational lensing causes the lensed images to become more extended, so this affects the detectability issue mentioned in Section \ref{subsec:detect}, where we assumed all images to be point sources. The most definite solution is to follow \citetalias{Collett15}: create mock QSO lenses using our predefined models, and test how many of them can be actually found in simulated observations. This is beyond the scope of this paper and deserves a paper of its own, and will not be discussed here further.

\subsubsection{A More Realistic Model}
For our calculations, QSO host galaxies are assumed to have singular isothermal spheroid mass distributions, which is a robust yet oversimplified model. Specifically, the two-image configuration predicted for these deflector mass models is critical for this work, in that the magnification of the dimmer image is used as the criterion for classifying QSO lenses. In the case of quad-like systems, the magnifications become more complicated. Once again, simulating mock QSO lenses will deliver more accurate results.

\section{Summary} \label{sec:summary}
In this paper, we introduce the HULQ project, which proposes to use QSO lenses to investigate the co-evolution of SMBHs and their host galaxies. To achieve this objective, an abundant sample of QSO lenses are required at various redshifts. We present the methodology and data to calculate the number of QSO lenses expected for various surveys. The main results are as follows.

\begin{enumerate}

\item The surface number density of QSO lenses is calculated as a function of the limiting magnitude of the imaging survey. Currently available surveys, such as PS1/3$\pi$ and publicly released HSC/Wide data are expected to provide $\sim 300$ QSO lenses, and this number will be augmented by at least one order of magnitude within the next decade. This justifies the feasibility of the HULQ project: discovering a statistically sufficient number of QSO lenses at various redshifts, with a significant portion at higher redshifts than currently known samples ($\zd \lesssim 0.5$).

\item The results above are verified by comparison with several studies regarding gravitational lenses in general. In particular, PDFs for several properties of QSO lenses are largely identical to those for gravitational lenses with normal galaxies as deflectors given in the literature. The expected numbers of QSO lenses for various surveys are also in line with those of galaxy lenses when assuming a $\sim 2.5$-dex difference between the two, which corresponds roughly to the number ratio of the two deflector populations at low redshifts (one in $\sim 300$ galaxies host QSOs). This implies that on average, the probabilities of normal and QSO host galaxies to be deflectors to background sources should be similar.

\begin{itemize}
\item In addition, our calculations were modified to be applicable to spectroscopic data. These results also agree well with those given in previous spectroscopic searches for QSO lenses, thus supporting our methods.
\end{itemize}

\item The effects of the evolution of the $\mbh - \sigma_*$ relation is discussed in detail. The most extreme evolution of the relation decreases the number of QSO lenses by a factor of $< 5$, so the HULQ project is still feasible. Studying the QSO lens sample in detail will allow us to determine the direction and amplitude of the redshift evolution of the relation in many approaches.

\item We simulated the effects of the Poisson noise of the QSO flux on the detectability of lensed images. Based on these simulations, we propose that QSO lenses with $\rein$ smaller than $\sim 1.7$ times the seeing of a survey cannot be detected, thus decreasing the number of \textit{detectable} QSO lenses; for HSC/Wide DR1, this factor is $\sim 20\%$. This demonstrates the importance of high-resolution imaging surveys, and underlines the significance of space-based surveys, both in discovering more QSO lenses and probing the low-mass QSO host galaxies.

\item We discuss various factors that may affect our results. Uncertainties in the derivation of the VDF do not change our results significantly ($\lesssim 0.2$ dex). The most critical factor is the over-simplification of the lens system design. The most direct solution is to create mock QSO lenses with more realistic models, and measure their detectability more accurately.

\end{enumerate}

To sum up, the future of the HULQ project is promising. Even when many factors are considered, the QSO lens sample discoverable with imaging data is expected to be abundant, especially with the advent of surveys conducted with space-based telescopes. This will provide us with a new method of studying the co-evolution of galaxies and their central SMBHs at high redshifts. 

\acknowledgments

We thank Yiseul Jeon and Yongjung Kim for useful discussions.

This work is supported by the National Research Foundation of Korea (NRF) grant, No. 2020R1A2C3011091, funded by the Korea government (MSIT).

The Hyper Suprime-Cam (HSC) collaboration includes the astronomical communities of Japan and Taiwan, and Princeton University. The HSC instrumentation and software were developed by the National Astronomical Observatory of Japan (NAOJ), the Kavli Institute for the Physics and Mathematics of the Universe (Kavli IPMU), the University of Tokyo, the High Energy Accelerator Research Organization (KEK), the Academia Sinica Institute for Astronomy and Astrophysics in Taiwan (ASIAA), and Princeton University. Funding was contributed by the FIRST program from Japanese Cabinet Office, the Ministry of Education, Culture, Sports, Science and Technology (MEXT), the Japan Society for the Promotion of Science (JSPS), Japan Science and Technology Agency (JST), the Toray Science Foundation, NAOJ, Kavli IPMU, KEK, ASIAA, and Princeton University. This paper makes use of software developed for the Large Synoptic Survey Telescope. We thank the LSST Project for making their code available as free software at  http://dm.lsst.org. The Pan-STARRS1 Surveys (PS1) have been made possible through contributions of the Institute for Astronomy, the University of Hawaii, the Pan-STARRS Project Office, the Max-Planck Society and its participating institutes, the Max Planck Institute for Astronomy, Heidelberg and the Max Planck Institute for Extraterrestrial Physics, Garching, The Johns Hopkins University, Durham University, the University of Edinburgh, Queen's University Belfast, the Harvard-Smithsonian Center for Astrophysics, the Las Cumbres Observatory Global Telescope Network Incorporated, the National Central University of Taiwan, the Space Telescope Science Institute, the National Aeronautics and Space Administration under Grant No. NNX08AR22G issued through the Planetary Science Division of the NASA Science Mission Directorate, the National Science Foundation under Grant No. AST-1238877, the University of Maryland, and Eotvos Lorand University (ELTE) and the Los Alamos National Laboratory. Based [in part] on data collected at the Subaru Telescope and retrieved from the HSC data archive system, which is operated by Subaru Telescope and Astronomy Data Center at National Astronomical Observatory of Japan.

Based on observations made with the NASA/ESA Hubble Space Telescope, obtained from the data archive at the Space Telescope Science Institute. STScI is operated by the Association of Universities for Research in Astronomy, Inc. under NASA contract NAS 5-26555.

Funding for the Sloan Digital Sky Survey IV has been provided by the Alfred P. Sloan Foundation, the U.S. Department of Energy Office of Science, and the Participating Institutions. SDSS-IV acknowledges support and resources from the Center for High-Performance Computing at the University of Utah. The SDSS web site is www.sdss.org. SDSS-IV is managed by the Astrophysical Research Consortium for the Participating Institutions of the SDSS Collaboration including the Brazilian Participation Group, the Carnegie Institution for Science, Carnegie Mellon University, the Chilean Participation Group, the French Participation Group, Harvard-Smithsonian Center for Astrophysics, Instituto de Astrof\'isica de Canarias, The Johns Hopkins University, Kavli Institute for the Physics and Mathematics of the Universe (IPMU) / University of Tokyo, the Korean Participation Group, Lawrence Berkeley National Laboratory, Leibniz Institut f\"ur Astrophysik Potsdam (AIP),  Max-Planck-Institut f\"ur Astronomie (MPIA Heidelberg), Max-Planck-Institut f\"ur Astrophysik (MPA Garching), Max-Planck-Institut f\"ur Extraterrestrische Physik (MPE), National Astronomical Observatories of China, New Mexico State University, New York University, University of Notre Dame, Observat\'ario Nacional / MCTI, The Ohio State University, Pennsylvania State University, Shanghai Astronomical Observatory, United Kingdom Participation Group,Universidad Nacional Aut\'onoma de M\'exico, University of Arizona, University of Colorado Boulder, University of Oxford, University of Portsmouth, University of Utah, University of Virginia, University of Washington, University of Wisconsin, Vanderbilt University, and Yale University.

\vspace{5mm}

\software{SExtractor \citep{Bertin+96}
          }

\clearpage

\clearpage

\clearpage

\clearpage

\clearpage

\clearpage

\clearpage

\

\clearpage

\

\clearpage

\clearpage

\clearpage

\clearpage

\clearpage

\clearpage

\begin{figure}
\centering
\epsscale{0.8}
\plotone{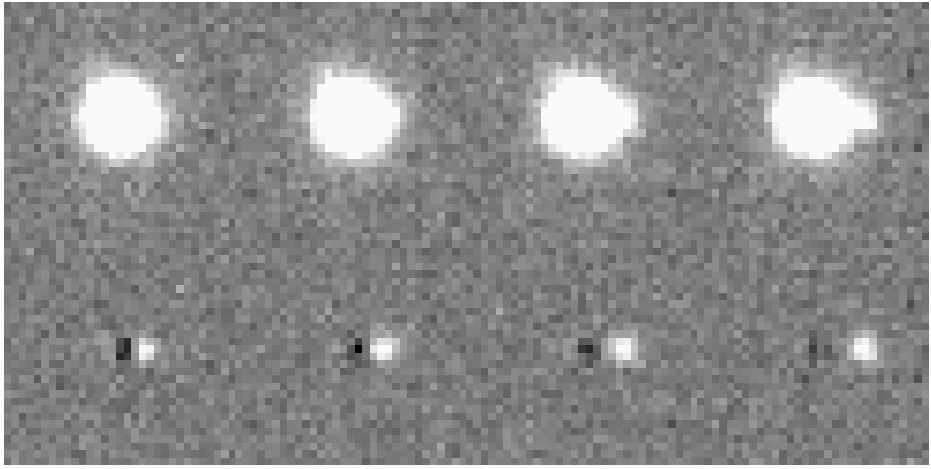}
\caption{(Top row) Simulated images of QSO lenses for four configurations as shown in Figure \ref{fig:detectability}, with the image conditions of HSC/Wide DR1.
(Bottom row) PSF-subtracted residual images for the four simulated images in the top row.
}
\label{fig:detect2}
\end{figure}

\clearpage

\begin{figure}
\centering
\epsscale{0.8}
\plotone{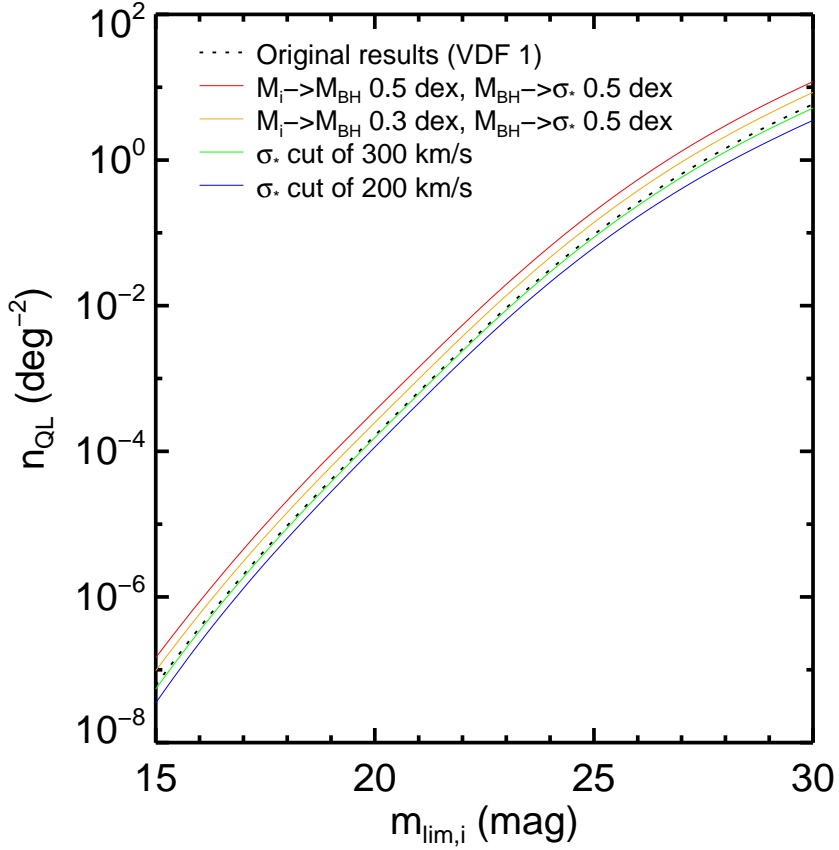}
\caption{$\nql$ versus $\mlim$ for various treatments to the VDF as discussed in Section \ref{subsec:uncer}. The dotted black line is for the original VDF, and is identical to the solid black line in Figure \ref{fig:fq1}. The red and yellow lines represent results for VDFs with increased scatters for the scaling relations; the yellow line with the scatter of the $\mbh - \sigma_*$ relation increased from 0.28 dex to 0.5 dex in the $\mbh$ direction, and the red line with the scatter of the $M_i - \mbh$ correlation increased from 0.3 dex to 0.5 dex, in addition. The green and blue lines represent results for VDFs with upper limits of 300 and 200 km s$^{-1}$, respectively, applied.
}
\label{fig:disc}
\end{figure}

\begin{deluxetable*}{clllc}
\tabletypesize{\scriptsize}
\tablecaption{Redshift Evolution of QSO and Galaxy Luminosity Function Parameters\label{tbl:QLF}}
\tablehead{
\colhead{} & \colhead{$\log \:\Phi^*$} & \colhead{$M_i^*/M_F^*$} & \colhead{$\alpha$} & \colhead{$\beta$}\\
\colhead{} & \colhead{(${\rm Mpc}^{-3}\:{\rm mag}^{-1}$)} & \colhead{(mag)} & \colhead{} & \colhead{}}
\startdata
QSO LF for $\zs < 2.2$	& $-5.96$ 						& $-22.85-2.5\times(1.241 \zs-0.249 \zs^2)$	& $-1.16$ & $-3.37$\\
QSO LF for $\zs > 2.2$ 	& $-5.96-0.689\times(\zs-2.2)$	& $-26.66-0.809\times(\zs-2.2)$				& $-1.16$ & $-3.37$\\
Galaxy LF for $\zs < 4$	& $-2.42-0.041\zs$ 				& $-21.74+0.20\zs$ 							& $-1.19-0.10\zs$ & -\\
Galaxy LF for $\zs > 4$	& $-2.58-0.32\times(\zs-4)$ & 	$-21.74+0.20\zs$ 							& $-1.19-0.10\zs$ & -\\
\enddata
\end{deluxetable*}

\begin{deluxetable*}{ccccccccccccccccc}
\tabletypesize{\scriptsize}
\tablecaption{Compilation of Galaxy Luminosity Functions\label{tbl:GLF1}}
\tablehead{
\colhead{Reference} & \colhead{Filter / Wavelength} & \colhead{Redshift} & \colhead{$\log \:\Phi^*$} & \colhead{$M_F^* (z=0)$} & \colhead{$\alpha$}\\
\colhead{} & \colhead{} & \colhead{} & \colhead{(${\rm Mpc}^{-3}\:{\rm mag}^{-1}$)} & \colhead{(mag)} & \colhead{}}
\startdata
\cite{Loveday+12}			& $r$ & 0.215 & $-2.34$ & $-21.55$ & $-$1.23\\
							& $i$ & 0.035 & $-2.40$ & $-21.65$ & $-$1.12\\
\cite{Ilbert+05}			& $B$ & 0.575 & $-$2.48 & $-$21.22 & $-$1.22\\
							& $B$ & 0.70 & $-$2.29 & $-$21.13 & $-$1.12\\
							& $V$ & 0.335 & $-$2.45 & $-$21.66 & $-$1.21\\
							& $V$ & 0.44 & $-$2.75 & $-$22.33 & $-$1.35\\
							& $R$ & 0.13 & $-$2.33 & $-$21.59 & $-$1.16\\
							& $R$ & 0.215 & $-$2.61 & $-$22.41 & $-$1.27\\
							& $I$ & 0.035 & $-$2.39 & $-$21.95 & $-$1.19\\
\cite{Marchesini+12}		& $V$	& 0.44 & $-$2.59 				& $-$21.76 				& $-$1.25	\\
\cite{Malkan+17}			& 1900 \AA	& $3.05$	& $-$2.73 			& $-$20.86 				& $-$1.78	\\
\cite{Sawicki+06}			& 1700 \AA	& $3.1$		& $-$2.77			& $-$20.90				& $-$1.43 \\
							& 1700 \AA	& $4$		& $-$3.07			& $-$21.00				& $-$1.26 \\
\cite{Reddy+09}				& 1700 \AA	& 3.05		& $-$2.77			& $-$20.97				& $-$1.73 \\
\cite{Parsa+16}				& 1500 \AA	& $3.8$		& $-$2.69			& $-$20.71				& $-$1.43	\\
\cite{Ono+18}				& $\sim1500$ \AA & $4$	& $-$2.52			& $-$20.63				& $-$1.57 \\
							& $\sim1500$ \AA & $5$	& $-$2.97			& $-$20.96				& $-$1.60 \\
							& $\sim1500$ \AA & $6$	& $-$3.27			& $-$20.91				& $-$1.87 \\
							& $\sim1500$ \AA & $7$	& $-$3.36			& $-$20.77				& $-$1.97 \\
\enddata
\end{deluxetable*}

\begin{deluxetable*}{llccccccccccccccccccccccccccc}
\tabletypesize{\scriptsize}
\tablecaption{Expected Number of QSO Lenses for Various Surveys\label{tbl:nqsolens}}
\tablehead{
\colhead{Survey} & \colhead{$m_{\rm lim,i}$$^1$} & \colhead{Area} & \colhead{Number of} & \colhead{PSF FWHM$^2$} & \colhead{Detectable number of }\\
\colhead{} & \colhead{} & \colhead{} & \colhead{QSO lenses} & \colhead{} & \colhead{QSO lenses}\\
\colhead{} & \colhead{(mag)} & \colhead{(deg$^2$)} & \colhead{} & \colhead{(\arcsec)} & \colhead{}}
\startdata
PS1/3$\pi$			& 23.1 & $\sim10000$ 	& 93	& 1.1	& 3.1\\
PS1/MDS				& 25.4$^3$ & 70			& 10	& 1.1$^3$	& 0.34\\
HSC/Wide 			& 26.2$^4$ & $\sim1400$ 	& 440	& 0.6	& 82\\
HSC/Wide DR1 		& 26.4 & $\sim100$ 		& 38	& 0.56	& 8.0\\
HSC/Deep 			& 27.1$^4$ & 26 			& 17	& 0.6	& 3.1\\
KiDS				& 24.2$^5$ & $\sim1500$ 	& 59	& 1.1	& 2.0\\
KiDS DR4			& 23.7$^5$ & $\sim470$		& 19	& 0.8	& 1.8\\
LSST (single visit)	& 24.0 & $\sim18000$	& 560	& 0.8	& 52\\
LSST (final)		& 26.8 & $\sim18000$	& 9700	& 0.8	& 900\\
\textit{Euclid}/Wide$^6$& 24.5$^7$ & $\sim15000$	& 740	& 0.23	& 480\\
\textit{Euclid}/Deep$^6$& 26.5$^7$ & $\sim40$		& 15	& 0.23	& 9.7\\
CSS-OS				& 25.9 & $\sim17500$	& 3800	& 0.15	& 3000\\
\textit{WFIRST}/WFIHLS$^8$& 26.7$^9$ & $\sim2200$	& 1000 & 0.18$^{10}$ & 720 \\
\enddata
\tablenotetext{1}{5-$\sigma$ limiting magnitude for point sources, unless noted otherwise.}
\tablenotetext{2}{$R_{\rm EE80}$ for space missions, with the exception of \textit{WFIRST}.}
\tablenotetext{3}{Obtained from \cite{Rest+14}.}
\tablenotetext{4}{Obtained from \cite{Aihara+18b}.}
\tablenotetext{5}{5-$\sigma$ limiting magnitude for 2$\arcsec$ apertures.}
\tablenotetext{6}{\textit{Euclid} is expected to observe in the optical wavelengths with the VIS instrument through a single wide filter, with wavelength coverages of 550-900nm, which is assumed to be equivalent to the $i$-filter here.}
\tablenotetext{7}{10-$\sigma$ limiting magnitude for extended sources.}
\tablenotetext{8}{\textit{WFIRST}/WFIHLS is not expected to conduct an optical survey; numbers are given assuming that a survey in the $i$-filter is undertaken. The details of the survey are for the $J$-filter portion of the survey.}
\tablenotetext{9}{5-$\sigma$ limiting magnitude, aperture not specified.}
\tablenotetext{10}{Obtained from \cite{Hounsell+18}, for the \textit{Z087}-filter.}
\end{deluxetable*}

\end{document}